\documentclass[letterpaper,twocolumn,10pt]{article}
\usepackage{usenix-2020-09}
\include{usenix-2020-09.sty}
\usepackage{balance}
\usepackage{appendix}
\usepackage[ruled,linesnumbered]{algorithm2e}
\usepackage{tikz}
\usepackage{amsmath}
\usepackage{algorithmic}

\usepackage{graphicx}
\usepackage{subcaption}
\usepackage{textcomp}
\usepackage{xcolor}
\usepackage{booktabs} 
\usepackage{multirow}
\usepackage{filecontents}
\usepackage{amsfonts}
\usepackage{mathtools}

\usepackage{ulem} 

\begin{document}
\title{\Large \bf MixDefense: A Defense-in-Depth Framework for Adversarial Example Detection Based on Statistical and Semantic Analysis}


\author{%
  Yijun Yang, Ruiyuan Gao, Yu Li, Qiuxia Lai, Qiang Xu* \\
  CUhk REliable Computing Laboratory (CURE)\\
  Department of Computer Science and Engirneering\\
  The Chinese University of Hong Kong\\
  {\{yjyang, rygao, yuli, qxlai, qxu\}@cse.cuhk.edu.hk} \\
}


 \maketitle

\begin{abstract}
Machine learning with deep neural networks (DNNs) has become one of the foundation techniques in many safety-critical systems, such as autonomous vehicles and medical diagnosis systems. DNN-based systems, however, are known to be vulnerable to adversarial examples (AEs) that are maliciously perturbed variants of legitimate inputs. While there has been a vast body of research to defend against AE attacks in the literature, the performances of existing defense techniques are still far from satisfactory, especially for adaptive attacks, wherein attackers are knowledgeable about the defense mechanisms and craft AEs accordingly.
In this work, we propose a multilayer defense-in-depth framework for AE detection, namely \textit{MixDefense}. For the first layer, we focus on those AEs with large perturbations. We propose to leverage the `noise' features extracted from the inputs to discover the \textit{statistical} difference between natural images and tampered ones for AE detection. For AEs with small perturbations, the inference result of such inputs would largely deviate from their \textit{semantic} information. Consequently, we propose a novel learning-based solution to model such contradictions for AE detection. Both layers are resilient to adaptive attacks because there do not exist gradient propagation paths for AE generation. Experimental results with various AE attack methods on image classification datasets show that the proposed MixDefense solution outperforms the existing AE detection techniques by a considerable margin. 
\end{abstract}

\section{Introduction}
\label{sec:introduction}
Deep neural networks (DNNs) have achieved unprecedented success in numerous long-standing machine learning tasks, and they are widely deployed in various safety-critical systems such as autonomous vehicles, healthcare systems, and cyber-security infrastructures.
In these applications, incorrect decisions or predictions could result in life risk or significant financial loss. Consequently, the safety and security of DNNs are of grave concern. 

\begin{figure*}[t]
    \centering
    \includegraphics[width=\textwidth]{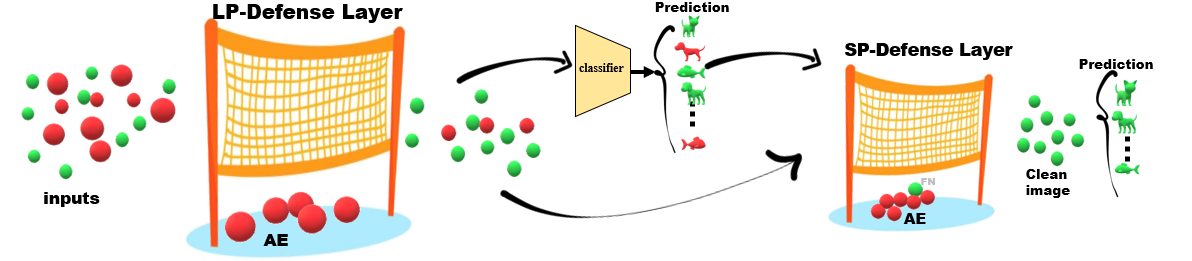}
    \caption{Overview of the proposed Defense-in-Depth framework \textit{MixDefense}. The first layer of defense is designed for AEs with large perturbation, and the second layer is for AEs with small perturbation. More layers can be added in between, if needed.}
    \label{fig:overview}
\end{figure*}

One of the primary threats to DNN-based systems is adversarial examples (AEs), which are maliciously perturbed inputs to fool the DNN model~\cite{szegedy2013intriguing}. Defending against AEs has drawn lots of research interests from both academia and industry~\cite{silva2020opportunities}. Existing AE defense techniques can be broadly categorized into three types: (i) \textit{AE-aware training}, including \textit{adversarial training} techniques that incorporate perturbed inputs into the training set (e.g.,~\cite{madry2017towards,pmlr-v97-pang19a}) and \textit{gradient masking/obfuscation} techniques that try to construct models with gradients that are difficult to use by attackers (e.g.,~\cite{Shan2020GottaCA,Lcuyer2019CertifiedRT}). While effectively mitigating AE threats, such solutions not only are computationally expensive but also cause an undesirable decrease in the prediction accuracy of legitimate inputs~\cite{pmlr-v119-raghunathan20a,tramer2020fundamental}. 
(ii). \textit{Input transformations}, which "purifies" the inputs before feeding them into the DNN (e.g.,~\cite{meng2017magnet,samangouei2018defense,Dziugaite2016ASO}). By doing so, carefully crafted adversarial perturbations are changed, thereby mitigating their abilities to attack.  
Due to the nature of this technique, there is an inherent tradeoff between tolerable perturbations and prediction accuracy for clean inputs. Consequently, such solutions are usually only effective for AEs with small perturbations, and most of them are not resistant to adaptive attacks.
 (iii). \textit{AE characterization}, which detects AEs by leveraging the instability of AEs (e.g.,~\cite{Grosse2017OnT, Song2018PixelDefendLG,Aigrain2019DetectingAE,Carrara2018AdversarialED}) or finding AEs as statistical outliers of all input samples (e.g.,~\cite{ma2018characterizing,kantaros2020visionguard,xu2018feature}).

 An ideal defense strategy for adversarial inputs should have the following features:
\begin{itemize}
    \item \textit{A broad scope of defense}, i.e., it should be able to defend against all sorts of AE attacks under a wide spectrum of perturbation strengths;
    \item \textit{Negligible impact on model accuracy with acceptable computational effort}, i.e., it should have a minimum impact on the prediction accuracy of legitimate inputs without incurring much training cost and inference overhead; 
    \item \textit{Robustness to adaptive attacks}, i.e., it should remain effective even if attackers are knowledgeable about the defense mechanism and craft AEs accordingly;
\end{itemize}
Existing solutions are still far from meeting the above criteria, despite there exist several ensemble solutions that combine multiple defense techniques (e.g.,~\cite{meng2017magnet,tramer2018ensemble, Cheval2018DEEPSECDE}).

In this work, we propose a novel defense-in-depth framework for adversarial examples, namely \textit{MixDefense}, which belongs to the "AE characterization" category. To achieve a broad scope of defense, we propose to use multiple \textit{attack-agnostic} layers to detect various kinds of AEs with different perturbation strengths, as shown in Fig.~\ref{fig:overview}. To be specific, we first use a \textit{LP-defense} layer for AEs with large perturbations, which leverages the 'noise' features extracted from the inputs to discover the \textit{statistical} difference between natural images and tampered ones. For AEs with small perturbations, the inference result of such inputs would largely deviate from their \textit{semantic} information. We use a learning-based \textit{SP-defense} layer to model the semantic distance between the input and its reconstruction conditioned on the predicted label for AE detection. As can be seen from Fig.~\ref{fig:overview}, the LP-Defense layer in MixDefense rejects those inputs that are deemed as AEs before feeding them into the DNN model and we strive to minimize overkills (i.e., legitimate inputs regarded as AEs) in this layer. The SP-defense layer is responsible for identifying the remaining AEs in a post-mortern manner. Such a defense strategy incurs little impact on model accuracy with small computational overhead. Moreover, as the final defense in the proposed MixDefense framework, the proposed SP-defense layer is resistant to adaptive attacks.

In summary, the contributions of this work are as follows:
\vspace{-3pt}
\begin{itemize}
\item We propose \textit{MixDefense}, a novel attack-agnostic defense-in-depth framework that can detect various kinds of AEs (including adaptive ones) while having little impact on the prediction accuracy of legitimate inputs with small inference overhead.

\vspace{-3pt}

\item For AEs with large perturbations, we propose a \textit{simple adversarial example catcher (SAEC)} technique as LP-defense layer (Section~\ref{sec:LP_defense}), which captures the intrinsic statistical difference between AEs and clean samples by measuring the `noise' level of the inputs.
\vspace{-3pt}

\item For AEs with small perturbation, we propose a learning-based soluton namely \textit{ContraNet} as \textit{SP-Defense} layer (Section~\ref{sec:SP_defense}).

Based on the contradiction between the semantic meaning of AEs and their predicted labels, this layer consists of a conditional generative model trained on clean images and their corresponding label pairs, and a deep metric module which learns a powerful similarity metric for measuring the distance between the input and the reconstructed image. 

\end{itemize}
\vspace{-3pt}

We verify the effectiveness of MixDefense on three popular image classification datasets: MNIST, CIFAR-10, GTSRB. Experimental results (Section~\ref{sec:results}) show that our method can defend against various AE attacks under a wide spectrum of perturbation strengths with little impact on the prediction accuracy for legitimate inputs, which outperforms the existing AE detection techniques by a considerable margin. Moreover, we perform case studies on adaptive attacks to MixDefense and show that our method is robust against them (Section~\ref{sec:adaptive_attack}).

\section{Background}
\label{sec:background}
A large amount of research has been dedicated to adversarial example attacks and defenses in the literature~\cite{silva2020opportunities}. In this section, we briefly introduce AE attacks and give more detailed discussions on existing AE defense solutions. 

\subsection{AE Attacks}
\label{sec:ae_attacks}
Adversarial example attacks can be broadly categorized into the following types:

\vspace{-3pt}
\begin{itemize}
    \item {\bfseries Gradient-based methods} generate adversarial perturbations according to the gradient direction of the loss function (e.g.,~\cite{fgsm,kurakin2016bim, madry2017towards, papernot2016limitations, Pang2020ATO, MoosaviDezfooli2016DeepFoolAS}). {\itshape Fast Gradient Sign Method (FGSM)}~\cite{fgsm} is the first AE attack that perturbs the inputs in a single step to achieve misclassification. Later attacks such as {\itshape Basic Iterative Method (BIM)}~\cite{kurakin2016bim}, {\itshape Projected Gradient Descent (PGD)}~\cite{madry2017towards}, and {\itshape Jacobian-based Saliency Map Attack (JSMA)}~\cite{papernot2016limitations} are more effective in terms of attack successful rate under a given perturbation strength constraint.
    \item {\bfseries Optimization-based methods} formulate AE designs as a constrained optimization problem and solve it accordingly. For example, {\itshape Carlini-Wagner (CW)}~\cite{cw2017} constructs the AEs by solving an optimization problem consisting of an L2-norm perturbation magnitude quantification term and a specific loss term. While more expensive than gradient-based AE attacks, to the best of our knowledge, it is one of the strongest AE attacks by far.
\vspace{-3pt}
    
    \item {\bfseries Transformation-based methods} conduct spatial transformations with slight geometric perturbations (e.g., {\itshape Spatially Transformed Attack}~\cite{xiao2018spatially}) or train a  generative model for AE transformation (e.g., {\itshape Adversarial Transformation Networks (ATN)}~\cite{baluja2017adversarial}). 

\vspace{-3pt}

\end{itemize}

\subsection{AE Defenses}
\label{sec:ae_defenses}
Numerous defense techniques are proposed in the literature to make DNN models more robust against adversarial perturbations~\cite{bulusu2020anomalous} and we categorize them as follows.

\setlength{\parskip}{0.5em}
\noindent
{\bfseries AE-aware training:} 
Adversarial training \cite{fgsm, kannan2018adversarial, ziang2018deepdefense, pmlrv97zhang19p, pmlr-v97-pang19a, tramer2018ensemble, madry2017towards} boosts model robustness by adding AEs into the training dataset so that their predictions can be corrected. Another type of AE-aware training strategy is to construct DNN models with gradients that are difficult to use by attackers, e.g., defensive distillation~\cite{papernot2016distillation}, randomized training~\cite{Liu2018TowardsRN}, adding gradient obfuscation layer~\cite{Liu2018TowardsRN,Lcuyer2019CertifiedRT}, and embedding trapdoors for AEs~\cite{Shan2020GottaCA}.

AE-aware training is relatively easy to perform when the target model is available and it is quite effective in terms of defense capability, especially for known AE attacks. However, due to the change of the target model, these techniques usually cause prediction accuracy loss~\cite{pmlr-v119-raghunathan20a,tramer2020fundamental}. Additionally, most of them are vulnerable to adaptive attacks.

\setlength{\parskip}{0.5em}
\noindent
{\bfseries Input transformations:}  Adversarial perturbations can be alleviated by input pre-processing, such as JPEG compression \cite{Dziugaite2016ASO, Das2018SHIELDFP}, bit-depth reduction \cite{Guo2018CounteringAI}, and randomized discretization \cite{Zhang2019DefendingAW, Xie2018MitigatingAE}. MagNet \cite{meng2017magnet} resorts to auto-encoders to learn the distribution of normal inputs, and then uses the trained auto-encoder to purify the input samples. 

\textit{Defense-GAN}~\cite{defense-gan-2018} is a novel defense mechanism that purifies the input into its closest GAN reconstruction. Similar idea is adopted by PixelDefend \cite{Song2018PixelDefendLG}, wherein a learned PixelCNN model is used to purify the inputs. 

The common limitation of input transformation-based defense techniques is that their defense capabilities are limited, especially when the adversarial perturbation is relatively large. Moreover, the prediction accuracy for legitimate inputs is affected because they are changed.

\setlength{\parskip}{0.5em}
\noindent
{\bfseries AE characterization:} Without changing the target DNN model and the inputs, AE characterization techniques find AEs directly.

One line of work regards AEs as statistical outliers. Some of them propose to train a binary classifier to learn the statistical differences between the distribution of clean data and that of the AEs \cite{Grosse2017OnT, Song2018PixelDefendLG, Metzen2017OnDA, Aigrain2019DetectingAE, Feinman2017DetectingAS, Carrara2018AdversarialED, ma2018characterizing,SAMP}.
In particular, Aigrain \textit{et. al.}~\cite{Aigrain2019DetectingAE}, Feinman \textit{et. al.}~\cite{Feinman2017DetectingAS}, and Metzen \textit{et. al.}~\cite{Metzen2017OnDA} use the intermediate representations from a pre-trained classifier to train the detector. \textit{Local Intrinsic Dimensionality (LID)}~\cite{ma2018characterizing} estimates a so-called LID value, i.e., the distance of the input and its neighbors, and detects AEs with large LID values. The above works train the detector with both legitimate and adversarial data in a supervised manner, and hence they are usually less effective for adaptive attacks. 

In contrast, some recent work proposes to use clean samples only to compute the probability of whether an input instance falls in the legitimate input distribution \cite{Zheng2018RobustDO, Ma2019NICDA, Grosse2017OnT, Song2018PixelDefendLG, Miller2019WhenNT}. Among them, \textit{PixelDefend}~\cite{Song2018PixelDefendLG} computes the probability with a generative model. 
Zheng \textit{et.al} model the  output distribution of the hidden neurons in a DNN classifier with a Gaussian Mixture Model~\cite{Zheng2018RobustDO}. 
Ma \textit{et. al.} propose to use the one-class SVM to extract the model in-variants (e.g. distribution of the neuron activation values) for anomaly detection~\cite{Ma2019NICDA}. Generally speaking, the above techniques are more effective for AEs with large perturbations as they are more easily to be identified as outliers.

\begin{figure*}[t!]
    \centering
    \includegraphics[width=0.8\linewidth]{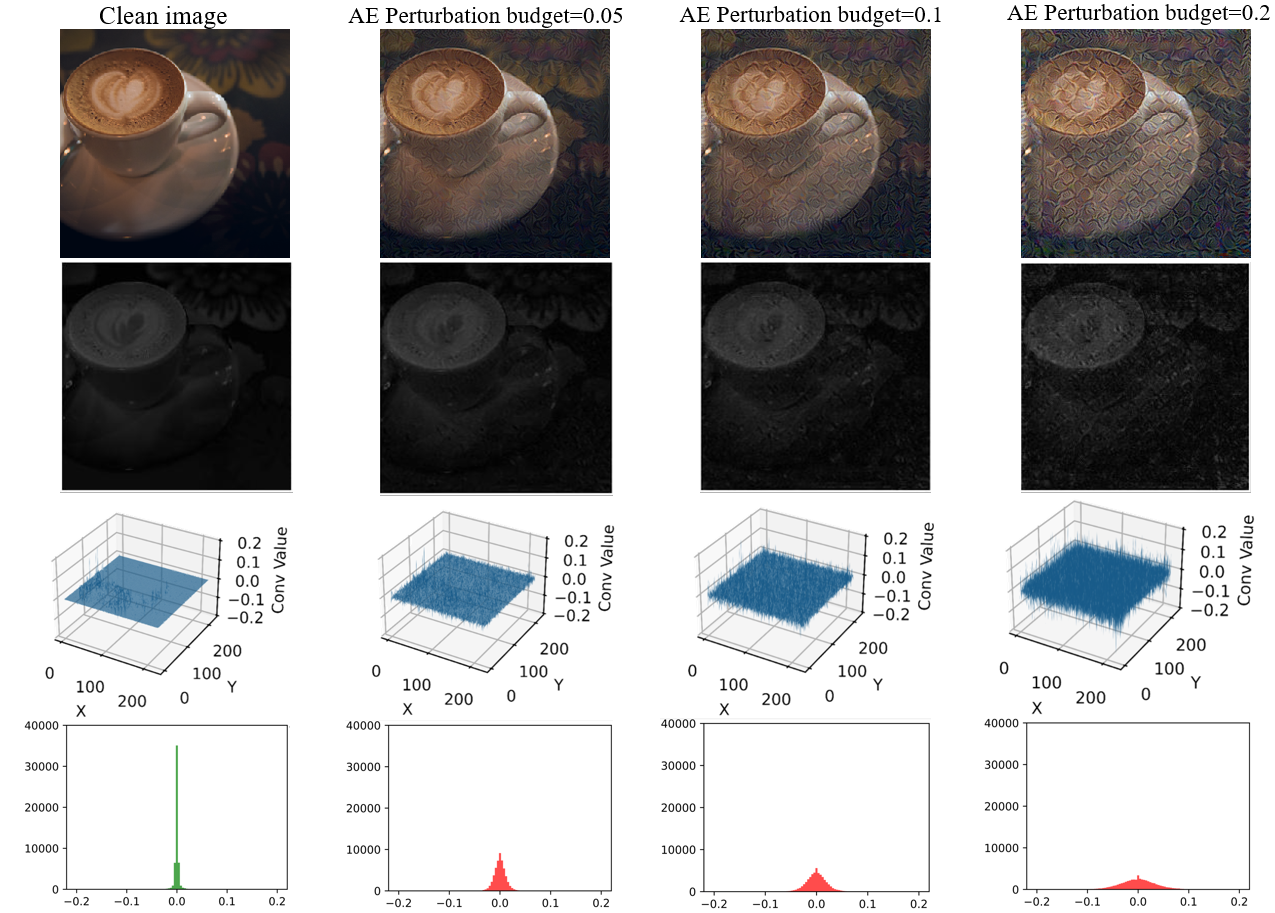}
    \put(-415,240){\fontsize{10pt}{10pt}\selectfont{(a)}}
    \put(-415,165){\fontsize{10pt}{10pt}\selectfont{(b)}}
    \put(-415,100){\fontsize{10pt}{10pt}\selectfont{(c)}}
    \put(-415,30){\fontsize{10pt}{10pt}\selectfont{(d)}}
    \caption{Given \textbf{(a) input images} with various perturbations, our proposed LP-Defense layer, SAEC, first obtains the \textbf{(b) pseudo saturation maps}, and then perform linear filtering to yield the \textbf{(c) noise contrast features}. We then calculate the noise level score based on the \textbf{(d) histogram} of the noise contrast features. We detect AEs by thresholding the noise level score.
    }
    \label{fig:large1}
\end{figure*}
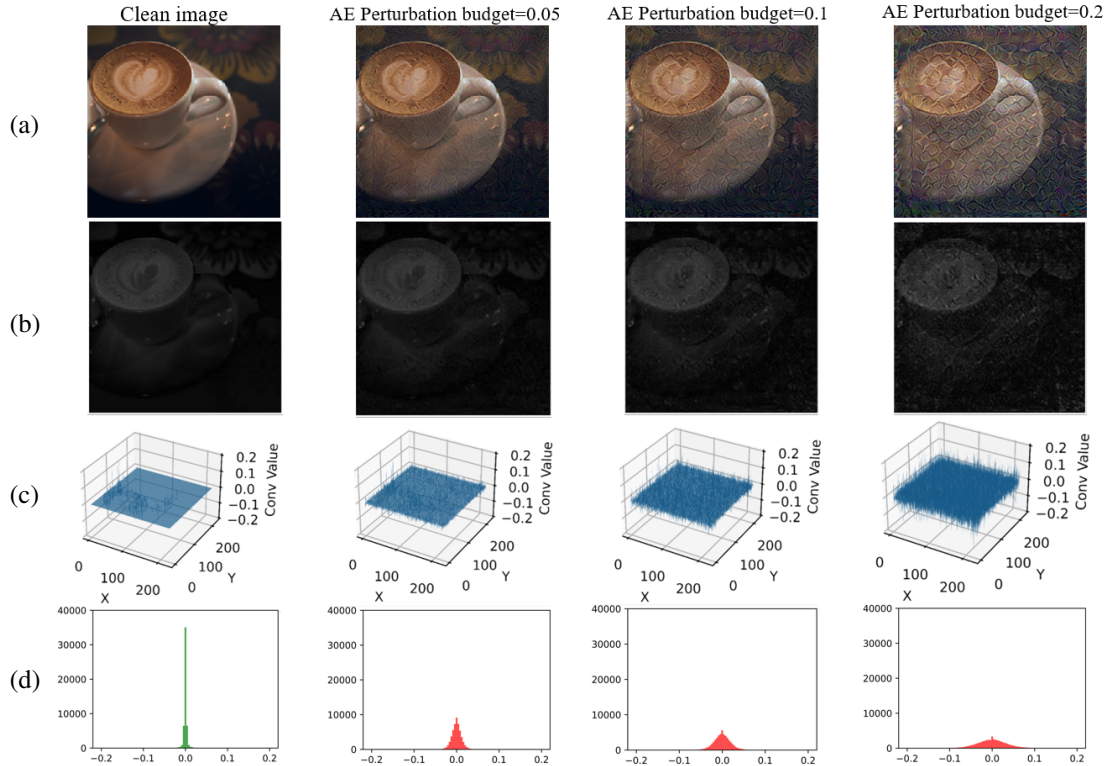

Another type of AE characterization work leverages the instability of AEs for detection~\cite{xu2018feature, kantaros2020visionguard, wang2019adversarial, liang2021}. \textit{Mutation Testing}~\cite{wang2019adversarial} performs model mutation (e.g., changing neuron activation values) and evaluates the prediction inconsistency for AE detection. While effective, it requires a large number of executions to run for each input sample, causing significant inference overheads. Liang \textit{et. al.}~\cite{liang2021} regard the adversarial perturbations as a kind of noise and implement two classical denoising techniques to mitigate their impact. Both the input image and its de-noised version are fed into the target classifier, and the input sample is regarded as an AE if their classification results are inconsistent. Similarly, Feature Squeezer (FS)~\cite{xu2018feature} and VisionGuard \cite{kantaros2020visionguard} propose to feed the input and its squeezed/compressed version to the DNN model and check the result difference for AE detection. While simple to deploy, these defense techniques provide moderate robustness enhancement for the target model, and they are also vulnerable to adaptive attacks.

Our proposed MixDefense framework belongs to the AE characterization category. In the following two sections, we detail the proposed detection technique against AEs with large perturbations and AEs with small perturbations, respectively.

\section{LP-Defense Layer}
\label{sec:LP_defense}

We model large adversarial perturbations as noise~\cite{liang2021} and propose a \textit{simple adversarial example catcher} (\textit{SAEC}) to serve as the LP-defense layer, which detects AEs by measuring the noise level of the input sample. However, it is challenging to determine whether the local pixel variations are due to the image texture, lighting variation, or the noise. 

To solve this problem, we first extract the noise contrast feature from the image, then calculate the noise level from the histogram of the feature map, and detect AEs by thresholding a noise level score.

\noindent\textbf{Noise contrast feature extraction.} 
To make the adversarial perturbations  more detectable, we propose to extract the noise contrast feature from the image. 
The feature extraction procedure consists of two steps: i) obtain a {\itshape pseudo-saturation map} and ii) perform linear filtering on the pseudo-saturation map. 

Inspired by the definition of the saturation in HVS space, we define the {\itshape pseudo-saturation map} as the $p$-norm of the image subtracted by pixel-wise mean intensity values: 

\begin{equation}
    \label{eq:saturationspace}
    \mathbf{x}_{stau} = \left (\sum_{c=1}^{3}(\mathbf{x}[:, :, c] - \mathbf{x}_{mean})^{p}  \right )^{1/p},
\end{equation}
where $\mathbf{x}\in\mathbb{R}^{W\times H\times 3}$ is the input RGB image, $\mathbf{x}_{mean}\in\mathbb{R}^{W\times H}$ is the mean intensity of each pixel, and $\mathbf{x}_{stau}\in\mathbb{R}^{W\times H}$ is the resulted pseudo-saturation map. 
As $p$ increases, the image content is suppressed, while the noises are exposed.
Therefore, it would be much easier to utilize statistical strategy to differentiate AEs.
In our experiments, we empirically set $p=8$.
Figure~\ref{fig:large1} (a) shows a batch of clean images together with their AEs in the RGB space. Figure~\ref{fig:large1} (b) displays their corresponding pseudo-saturation projections. 
As can be observed, the image content such as shape, color, and brightness are dramatically declined, while the noise component (salt and pepper-like points) is largely preserved,  which are exactly the adversarial perturbations.

\begin{figure*}
    \centering
    \includegraphics[width=\linewidth]{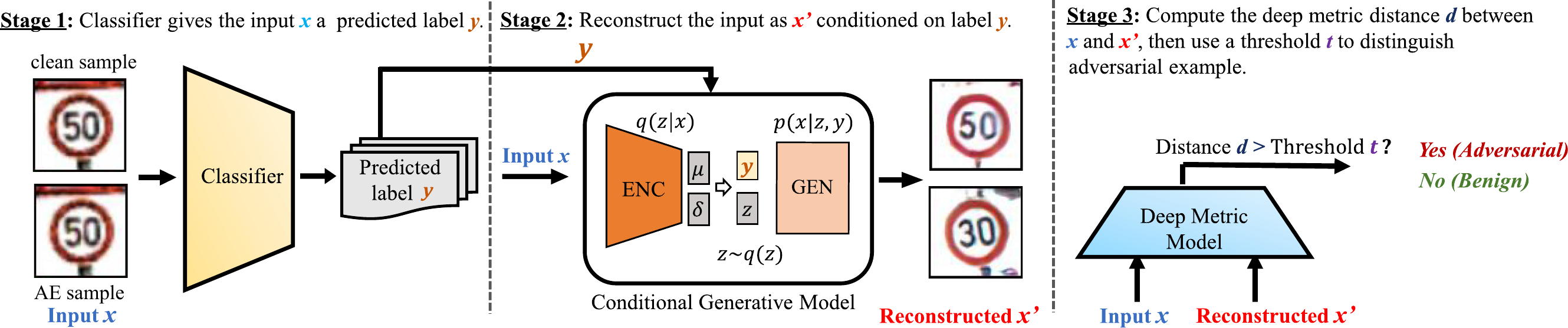}
    \caption{
    The overview of the proposed SP-Defense layer, ContraNet.
    During inference, in the $1$st stage, the target classifier assigns the input image $\mathbf{x}$ a predicted label $y$. In the $2$nd stage, we pass both the input image $\mathbf{x}$ and its predicted label $y$ to the conditional generative model, which reconstructs $\mathbf{x}$ based on $y$. After that, the distance between $\mathbf{x}$ and its reconstruction $\mathbf{x}'$ is measured by the trained deep metric model. If the distance is larger than the threshold $t$, the input instance is detected as an AE. }
    \label{fig:contranet}
\end{figure*}


 To further utilize the pseudo-saturation map $\mathbf{x}_{stau}$, we leverage a linear filter to augment the variances between the smoothed image content and the exposed perturbed regions. 
Such linear filters have been widely used in steganalysis of digital images~\cite{steg1,steg0,steg3,steg4,steg5}, and it has also been used for AE detection in~\cite{liu2019detection}. Different from~\cite{liu2019detection}, we apply the filters on the pseudo-saturation maps instead of on the natural images.
As shown in Figure~\ref{fig:large1} (c), the distinction between the noise and the image content $\mathbf{x}_{stau}$ is further amplified. Therefore, such linear filters can further help us differentiate AEs from clean samples.

\noindent \textbf{AE Detection.}
After extracting the noise contrast feature, we first obtain the histogram $\mathbf{H}$ of the feature, and then calculate the corresponding noise level scores. 
Exemplar histograms are shown in Figure~\ref{fig:large1} (d). Here, the $x$-axis represents the value of the noise contrast feature, and $y$-axis represents the number of pixels falling in that bin. The red and green bars represent the histograms of AEs and clean images, respectively. 
As can be observed, the values of noise contrast feature for clean images tend to concentrate around zero, while the values of AEs distribute among a wider range.

 Furthermore, to quantify such observations for utilization, we define the \textit{noise level score} (\textit{NL\_score}) as the variance of the histogram $\mathbf{H}$:

\begin{equation}
    \label{eq:score}
    \text{NL\_score} = \frac{1}{L} \sum_{i=1}^{L} (\mathbf{H}[i] - \mu)^{2},
\end{equation}

where $L$ is the number of bins in $\mathbf{H}$, $\mathbf{H}[i]$ denotes the value of the $i$-th bin, and $\mu=\frac{1}{L} \sum_{i=1}^{L} \mathbf{H}[i]$ is the mean value. 
 
Finally, we can detect AEs by directly thresholding the NL\_scores. The threshold value is determined in such a manner that clean images would not be mistakenly regarded as AEs. 

SAEC has several advantages: (i). it is attack-agnostic and hence can be applied to all kinds of AE attacks; (ii). it has a negligible computational cost.

As small perturbations do not yield statistically significant variations on the pixel value distribution of the natural images, many AEs (mostly with small perturbations) can bypass our LP-defense layer and they are detected by the SP-Defense layer, as detailed in the following section.

\section{SP-Defense Layer}
\label{sec:SP_defense}

For successful AEs with small perturbations, their inference results (with wrong labels) would naturally contradict with their semantic information observed by a human. Moreover, the smaller the perturbation, the more evident such contradiction is. Based on this observation, in this section, we present \textit{ContraNet}, a learning-based AE detection technique to serve as the \textit{SP-Defense} layer in our MixDefense framework.

ContraNet consists of a class-conditional generative model trained on the clean image and label pairs, and a deep metric module which learns a strong similarity metric for measuring the distance between the input and the reconstructed image.
Figure \ref{fig:contranet} presents the workflow of the proposed solution.
Given a DNN-based target classifier $f(\mathbf{x};\theta)$, we utilize a class-conditional generative model $g(\mathbf{x},y;\theta)$ (Section~\ref{sec:cgan}) to learn the joint distribution between input images and their labels. To be specific, we train $g(\mathbf{x},y;\theta)$ with clean training samples $\mathbf{x} \in \mathcal{X}_{\text{train}}$ to reconstruct the inputs conditioned on the corresponding class labels $y \in \mathcal{Y}_{\text{train}}$. During inference, for a given input $\mathbf{x} \in \mathcal{X}_{\text{test}}$, we reconstruct it (in Stage 2) with $g(\mathbf{x},y;\theta)$ conditioned on its predicted label $y' = f(\mathbf{x}; \theta)$. Generally speaking, for a legitimate input with correct prediction, its reconstruction $\mathbf{x}'=g(\mathbf{x},y';\theta)$ would be similar to itself $\mathbf{x}$; otherwise, if the input is an adversarial example that is obviously not associated with the predicted label, the reconstructed sample would not be faithful, because the class-conditional reconstruction model has not seen such kind of input and label pair during training. Next, in Stage 3, by measuring the distance between the input image $\mathbf{x}$ and its reconstructed counterpart $\mathbf{x}'$, we can distinguish the adversarial example from the benign ones. It is worth noting that, we need to measure the distance from a semantic perspective. Therefore, instead of using rule-based distances (e.g., $L_p$ or Cosine distance), which is notoriously poor in measuring distances in high dimensional space, we propose to use a learned deep metric model (Section~\ref{sec:deep_metric}) for distance measurement.

\subsection{Class-Conditional Generative Model} 
\label{sec:cgan}

\noindent \textbf{Framework overview.}
One straightforward way to realize the class conditional generation is to use an existing conditional generative adversarial network (cGAN \cite{miyato2018cgans}), which can generate images that are highly associated with the given condition. 
However, as existing cGAN techniques generate images from a randomly sampled latent vector $\mathbf{z}$ from Gaussian distribution, the generated image is not directly related to the input $\mathbf{x}$. Consequently, they can be very different from $\mathbf{x}$ even under the correct condition label (e.g., the input can be "red bird flying in the blue sky" while the generated image can be "yellow bird standing on the green tree").

By doing so, given the correct condition, the generated image would be a faithful reconstruction of the input image instead of a random image of this category.

\begin{figure}[t]
    \centering
    \includegraphics[width=\linewidth]{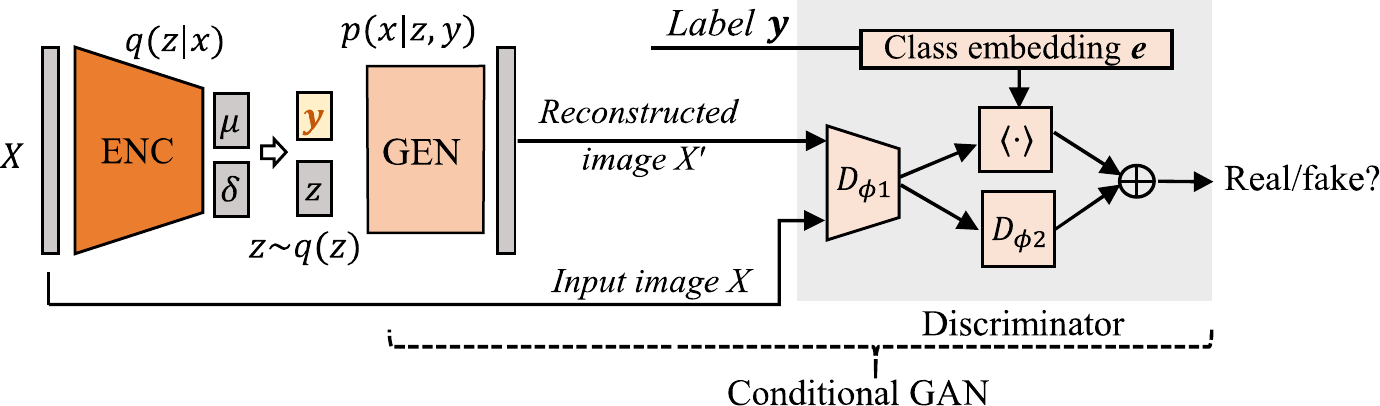}
    \caption{The revised cGAN in ContraNet. It mainly contains three parts, the encoder, the generator, and the discriminator. 
    Especially, the discriminator 1) helps the generator to improve its reconstruction quality; 2) makes the generator not ignore the class condition. Please note that the discriminator is only used during training. 
    See Section~\ref{sec:cgan} for details.}
    \label{fig:cvae}
\end{figure}

\begin{algorithm}[t]
\caption{\small Training of the revised cGAN.
}
\begin{algorithmic}[1]
\label{alg:train_CGM}
\FOR{number of training iterations}
  \FOR{$k$ steps}
    \STATE{$\bullet$ Sample a mini-batch of $m$ examples 
    $\{(\mathbf{x}^{(i)}, y^{(i)})\}_{i=1}^m$
    from $(\mathcal{X}_{\text{train}}, \mathcal{Y}_{\text{train}})$.}
        \STATE{$\bullet$ Update the generator by gradient descent:
            \[
                \nabla_{\theta_g} \frac{1}{m} \sum_{i=1}^m
                \log \left(1-D(G(E(\mathbf{x}^{(i)}), y^{(i)}))\right)
                .
            \]}
        \[
            \nabla_{\theta_g} \frac{1}{m} \sum_{i=1}^m \left[
            Dist \left(G(E(\mathbf{x}^{(i)}),  y^{(i)}),  \mathbf{x}^{(i)}\right)
            \right]. \]

    \STATE{$\bullet$ Update the encoder by gradient descent:
        \[
            \nabla_{\theta_e} \frac{1}{m} \sum_{i=1}^m \left[
            Dist \left(G(E(\mathbf{x}^{(i)}),  y^{(i)}),  \mathbf{x}^{(i)}\right)
            \right]. \]
        \[ \nabla_{\theta_e} \frac{1}{m} \sum_{i=1}^m \sum_{j=1}^J \left[
            \sigma_j^{(i)} + (\mu_j^{(i)})^2 - 1 - (\log \sigma_j^{(i)})
            \right].
        \]}

    \STATE{$\bullet$ Update the discriminator by gradient descent:
        \[
            \nabla_{\theta_d} \frac{1}{m} \sum_{i=1}^m \left[ 
            \log D(\mathbf{x}^{(i)},  y^{(i)})\right]
            \]
        \[
            \nabla_{\theta_d} \frac{1}{m} \sum_{i=1}^m \left[\log \left(1-D(G(E(\mathbf{x}^{(i)}),  y^{(i)}),  y^{(i)})\right)
            \right].
        \]}
   \ENDFOR

  \ENDFOR
  \\
    \textit{The \textit{Dist} function can be any distance measurement functions. In our experiment, we use the $L_2$ and $SSIM$.}
\end{algorithmic}
\end{algorithm}

Figure \ref{fig:cvae} shows our cGAN architecture. The model contains an encoder $q(\mathbf{z}|\mathbf{x})$, a generator $p(\mathbf{x}|\mathbf{z},y)$, and a discriminator $[D_{\phi_1}, D_{\phi_2}]$. All these modules are parameterized by feed-forward neural networks. Given an input $\mathbf{x}$, the encoder encodes the input image into a latent vector $\mathbf{z}$. Later, the latent vector and the condition $y$ are feed to the generator to generate image of class $y$.
At last, the conditional discriminator is responsible for discriminating the real image $\mathbf{x}$ and the generated fake image $\mathbf{x}'$ to improve generation quality.

To make sure that the condition is not ignored by the generator, we insert the condition information into both the generator and the discriminator following the practice in \cite{miyato2018cgans}. First, we insert one Class Batch Normalization (CBN) layer into each layer of the generator. Second, we use a projection layer to project the class embedding into the discriminator output. The right part of Figure \ref{fig:cvae} shows the discriminator details. First, the label is processed by an embedding layer to obtain its class embedding $e$. Then, we apply an inner product between $e$ and the extracted feature $D_{\phi_1(\mathbf{x})}$. Later, the two branches are added together to produce the output result.

\noindent \textbf{Training process.}
The training of the three components in the revised cGAN is detailed as follows.
Given the dataset with input images and the associate ground-truth labels $(\mathbf{x}, y) \in (\mathcal{X}_{\text{train}}, \mathcal{Y}_{\text{train}})$, the encoder is optimized in a way that the reconstructed $\mathbf{x}'$ is similar to the input $\mathbf{x}$ conditioned on the label $y$.  Besides, we still maintain the distribution of the latent vector $\mathbf{x}$ to be Gaussian distribution. Therefore, the objective of the encoder model is formally defined as:
\begin{multline}
\label{eq:encoder_loss}
     \mathcal{L}(E) = E_{p(\mathbf x)} D_{KL}(q (\mathbf z|\mathbf x) || N(0, \mathbf I)) \\ - E_{p(\mathbf x)}E_{p(\mathbf z|\mathbf x)} [\log p(\mathbf x|\mathbf z, y)],
\end{multline}
where the first term is the KL divergence (between $\mathbf{z}$ and Gaussian distribution) loss, and the second term denotes the reconstruction loss. 

For the generator, except the reconstruction loss, we also expect the generated image is able to cheat the discriminator. Thus, if we use $d(u|\mathbf{x},y)$ to denote the discriminator, the loss function of the generator is defined by:
\begin{multline}
\label{eq:gen_loss}
    \mathcal{L}(G) = - E_{p(\mathbf x)}E_{p(\mathbf z|\mathbf x)} [\log p(\mathbf x|\mathbf z, y)] \\ - E_{p(y)} E_{p(\mathbf x|y)}[\log(d(1|\mathbf x, y))].
\end{multline} 

At last, as the discriminator is trained to distinguish the real input image and the generated one, the adversarial loss for the discriminator is given by:
\begin{multline}
\label{eq:dis_loss}
	\mathcal{L}(D) = -E_{p(y)} E_{p(\mathbf x|y)}[\log(d(1|\mathbf x, y))] \\
	- E_{q(\mathbf z|\mathbf x)}E_{p(\mathbf x|\mathbf z, y)}[\log (d(0|\mathbf x, y))].
\end{multline}

Algorithm \ref{alg:train_CGM} shows the mini-batch realization of the above training objectives ($\mathcal{L}(G)$, $\mathcal{L}(E)$, and $\mathcal{L}(D)$). First, in each training iteration, we split the training into $k$ mini-batches and each mini-batch contains of $m$ samples. In each mini-batch, we calculate the respective gradient w.r.t. the model parameters with the loss function $\mathcal{L}(G)$, $\mathcal{L}(E)$, and $\mathcal{L}(D)$. Then, we update the models towards the gradient descending direction. The training iteration is repeated for a pre-defined number of training iterations.

After the encoder, the generator, and the discriminator are trained, only the encoder and the generator are used for inference. The discriminator is only an auxiliary model to help the revised cGAN consider the label information.

\subsection{Deep Metric Learning} 
\label{sec:deep_metric}

Directly applying traditional similarity metrics (e.g., $L_p$ norm, or Cosine similarity) to measure the distance between the input and the reconstructed image often results in suboptimal detection results. To tackle this problem, we propose to use deep metric learning to learn a more powerful similarity metric for AE detection.

\begin{table}[t]\centering
  \caption{Computation and  storage cost of ContraNet}
  \label{tab:computiation}
  \begin{tabular}{ccl}
    \toprule
    Component & Params(M) & FLOPs(G) \\
    \midrule
    Encoder &3.81& 0.11 \\
    Generator & 4.3 & 1.69\\
   Deep Metric Model& 2.24 &0.03 \\
   Total & 10.35&1.83\\
    \bottomrule
 
  \end{tabular}

\end{table}

To be specific, we employ a {\itshape Triplet network}\cite{triplet} as our deep metric module for learning the similarity metric. 
The deep metric module consists of three instances of the same feedforward network $M$ with shared parameters. 
This module has three types of inputs, namely, the anchor $\mathbf{x}$, the positive sample $\mathbf{x}^+$, and the negative sample $\mathbf{x}^-$. 
Here, we use clean images as anchors and obtain the positive and negative samples by feeding the clean images into the revised cGAN. When accompanied with correct labels, the output reconstructions are marked as positive samples; otherwise, they are marked as negative samples. 
These three kinds of samples are fed into the deep metric module to get their respective embeddings.

For training, we utilize {\itshape Triplet Margin Loss}\cite{tripletloss} as the loss function, as depicted in Equation\eqref{tripletloss}.

\begin{multline}
    \mathcal{L}_{triplet} =
    max\{d(M(x), M(x^{+}))\\-d(M(x),M(x^{-}))+margin,0 \},
    \label{tripletloss}
\end{multline}
where the $margin$ is a positive number which further enlarging the distance among dissimilar samples and $d$ demonstrates regular distance metric, e.g. Euclidean distance, consine distance which measures the distance between embedding vectors obtained from $M$.

To improve the performance, all possible triplet combinations are used in loss calculation according to the labels. 
We also resort to hard sample miner\cite{miner, metric0} during training.

\begin{table}[t]
  \caption{Datasets and Target Classifier}
  \label{tab:DataSets}
  \begin{tabular}{cccl}
    \toprule
    Dataset & No. of Classes & Classifier Type & Accuracy \\
    \midrule
    \texttt{MNIST\cite{mnist}} & 10& LeNet& 98.6\% \\
    \texttt{CIFAR10\cite{krizhevsky2014cifar}}& 10 & DenseNet169& 94.3\% \\
    \texttt{GTSRB\cite{GTSRB}}& 43 &ResNet34& 97.8\% \\
    \bottomrule
 
  \end{tabular}
     
      \footnotesize\textit{Accuracy:} refers to the classifier's accuracy on clean images.
\end{table}

One distinguished advantage of ContraNet is its robustness to adaptive attacks. First, the generative model itself is more robust than discriminative model under adversarial attack~\cite{kos2018adversarial}. Also, the target classifier and our detection mechanism have separate gradient propagation paths. The only connection between them is the predicted label produced by a non-differentiable \texttt{argmax} operation. Therefore, the attacker cannot implement an end-to-end gradient-based adaptive attack. Moreover, similar to the SAEC technique used in the LP-defense layer, ContraNet is also an attack-agnostic plug-and-play defense solution that requires the training dataset only. In Table~\ref{tab:computiation}, we show the number of parameters and the number of GFLOPS of ContraNet for CIFA10 dataset. As can be observed, the size and inference cost of ContraNet is moderate, considering the fact that classifiers used in safety-critical applications are usually quite large to achieve extremely high accuracy.

\section{Experimental Results}
\label{sec:results}

In this section, we first introduce the experimental setting in Section~\ref{sec:eval_metrics}.
Then, we report the performance of MixDefense from three aspects: the overall performance of MixDefense compared with three state-of-the-art detection-based defense methods (Section~\ref{sec:exp_mix}), the performance of the proposed LP-Defense layer SAEC (Section~\ref{sec:exp_lp}), and the performance of the proposed SP-Defense layer ContraNet (Section~\ref{sec:exp_sp}).

\subsection{Experimental Settings}
\label{sec:eval_metrics}

\noindent
We evaluate our MixDefense on three popular datasets: MNIST\cite{mnist}, CIFAR10\cite{krizhevsky2014cifar}, and GTSRB\cite{GTSRB}. 
The datasets along with the target classifiers are summarized in Table~\ref{tab:DataSets}.

\noindent\textbf{Evaluation metrics.} 
To evaluate the performance of AE defenses~\cite{carlini2019evaluating}, we use the two metrics described in \cite{deepsec} and \cite{meng2017magnet}.

The first metric is the detection accuracy of the detector given a half-and-half mixture of clean images and AEs as inputs. We treat clean images as positive samples, and AEs as negative samples. The accuracy of the detector is calculated as:
\begin{equation}
    Acc_{\text{Detector}} = \frac{TP+TN}{TP+TN+FP+FN}, 
    \label{accDetector}
\end{equation}
where $TP, TN, FP, FN$ denotes true positive, true negative, false positive and false negative, respectively.

The second metric is the overall accuracy of a \textit{robust classifier} (\textit{RC}), i.e., a classifier equipped with an AE detector, which is calculated as the percentage of AEs that are either detected by the detector, or correctly classified by the classifier:
\begin{equation}
    Acc_{\text{RC}} = \frac{\#Detected + \#Correct}{\# AEs}. 
    \label{accrobust}
\end{equation}
Here, $\#AEs$ is the total number of AEs (including the ones that do not attack successfully), $\#Detected$ is the number of AEs detected by the detector, and $\#Correct$ is the number of AEs that are correctly classified by the target classifier.

\noindent\textbf{Perturbation budget.} The perturbation budget $\epsilon$ measures the allowed differences between the clean sample $\mathbf{x}$ and the produced AE $\mathbf{x}^{adv}$ :
\begin{equation}
\label{pnorm}
    \left \|\mathbf{x}^{adv}-\mathbf{x}  \right \|_{p} \leq  \epsilon, 
\end{equation}
where $\|\cdot \|_{p}$ denotes $p$-norm.The larger the perturbation budget $\epsilon$ is, the stronger the produced AEs are.

\noindent\textbf{Compared defense methods.}
We compare MixDefense with two competing detection-based defenses FS~\cite{xu2018feature} and MagNet~\cite{meng2017magnet} on the accuracy of detector ($Acc_{\text{Detector}}$); and with FS~\cite{xu2018feature}, MagNet~\cite{meng2017magnet} and one of the most strongest training-based defense known today, Madry~\cite{madry2017towards}, on the accuracy of robust classifier ($Acc_{\text{RC}}$). 
The selected methods are representative AE defense methods from the three categories surveyed in Section~\ref{sec:ae_defenses}, where Madry boosts model robustness through adversarial training, MagNet alleviates the impact of AEs through input transformation, and FS detects AEs based on the prediction inconsistency.
All three methods are implemented based on their open-source official code. Note that, since the adversarial training is time-consuming, we only test Mardry on MNIST and CIFAR10, and use the pre-trained models provided by the authors.

We also use a large perturbation budget for the comparison experiments to show the worst-case performance of the defense methods. 
For defense models realized using Pytorch, we use the implementations of FGSM, FGM, BIM, C\&W, and DeepFool from the foolbox library~\cite{foolbox}. 
For those realized with TensorFlow, \textit{e.g.}, MagNet, we resort to cleverhans~\cite{cleverhans} to implemented the attacks. Table \ref{tab:Attack Methods} summarizes the attacks we employed. 

\subsection{Results of MixDefense}
\label{sec:exp_mix}
This subsection is concerned with MixDefense's overall performance against various attacks under a wide range of adversarial perturbations. 
Here, we only report results of attacks using $L_{2}$ norm as distance measurement, and put results of $L_{\infty}$ in Appendix due to limited space.

\begin{table}[t]
  \caption{Attack Methods}
  \label{tab:Attack Methods}
  \begin{tabular}{cccl}
    \toprule
    Attack Method & Knowledge & Goals & Distance \\
    \midrule
    \texttt{FGSM\cite{fgsm}} & whitebox& untarget&  $L_{\infty}$ \\
    \texttt{FGM\cite{fgsm}} & whitebox&untarget & $L_2$\\
    \texttt{BIM\cite{kurakin2016bim}}& whitebox & untarget& $L_2, L_{\infty}$ \\
    \texttt{C\&W\cite{cw2017}}& whitebox &untarget& $L_2$ \\
    \texttt{DeepFool\cite{moosavi2016deepfool}}& whitebox & untarget & $ L_2, L_{\infty}$\\
    \bottomrule
 
  \end{tabular}
     
\end{table}

\begin{figure*}[t]
    \centering
    \includegraphics[width=\linewidth]{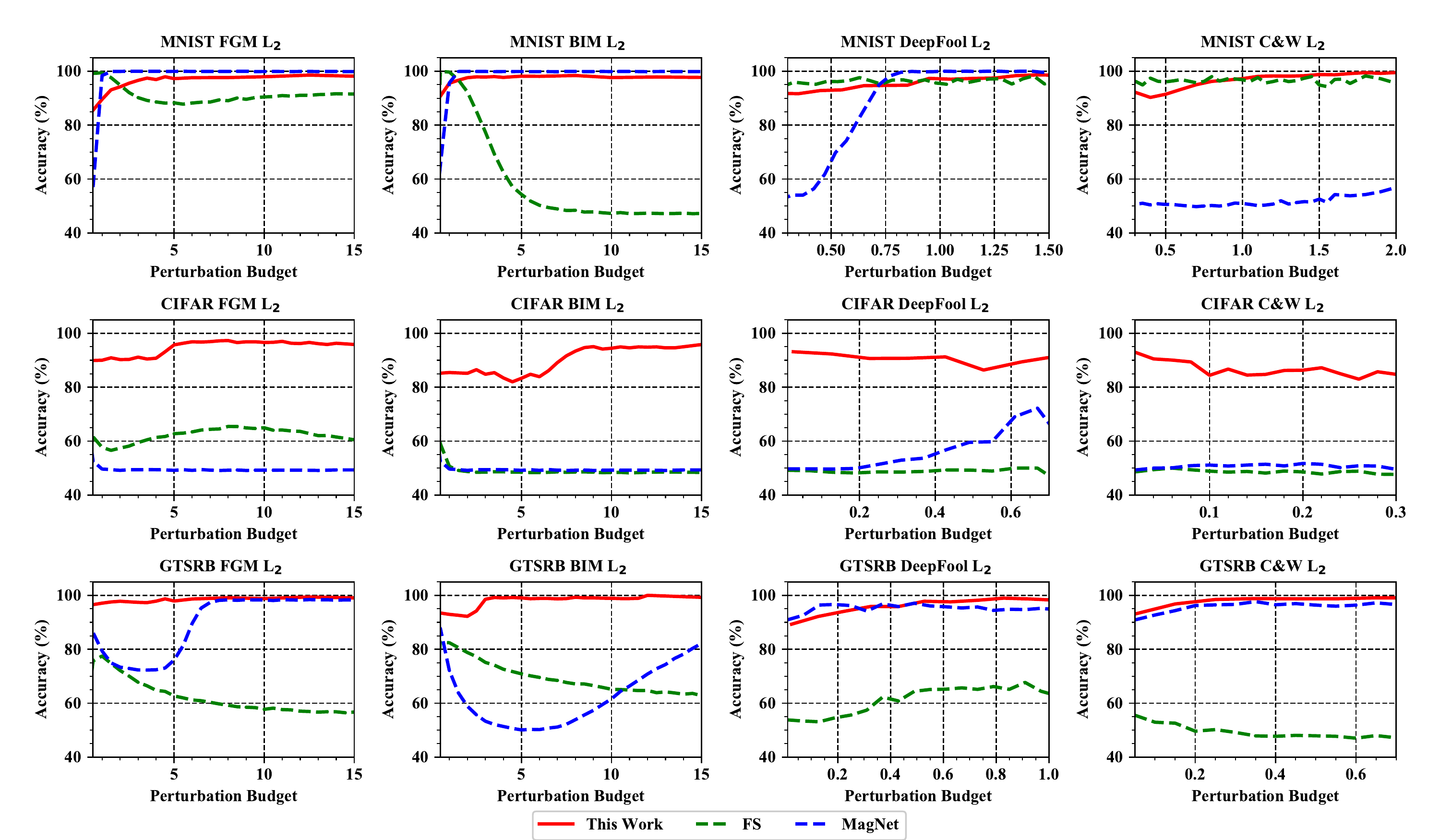}
    \caption{{\itshape{$Acc_{Detector}$ vs. perturbation budget} } curves  under $L_{2}$  norm across 3 datasets.  The column presents attack methods, and the row shows the type of dataset. The red solid line refers to our {\itshape{MixDefense}}, while the green dash line and blue dash line display the performance of {\itshape{FS}} and {\itshape{MagNet}} respectively.}
    \label{fig:detectorACCL2}
\end{figure*}

\noindent
{\bfseries Accuracy of Detector.} 
Figure~\ref{fig:detectorACCL2} summarizes the $Acc_{\text{Detector}}$ of MagNet, FS, and our MixDefense under different attacks with various perturbation budgets.  Each accuracy curve is fitted over 30 ascending adversarial perturbation budgets. For FGSM, FGM and BIM attack, each point is obtained leveraging around 1000 randomly sampled images from the test set, containing half clean images and half AEs. Note that, when the perturbation budget is too small, it is hard for some attack method to fool the classifier. If that happens, we collect all possible success AEs to perform the experiments. Once the collected AE number is less than 100, we will simply crop the corresponding perturbation budgets, that is the reason why each panel's perturbation scales are not strictly equal. As for DeepFool and C\&W,  there are no explicit attack parameters to control the adversary perturbation budgets. We call the attack once, then sort the generated adversarial samples by their perturbation budgets and plot the accuracy curve based on it.   Additionally, the threshold is fixed for each dataset, that is to say, we use the same threshold against all kinds of attacks. Unless specified otherwise, the above setting is used for the following experiments about accuracy curves.

As can be observed, MixDefense can defend both AEs with large perturbation and AEs with small perturbation. This is because MixDefense consists of multiple defense layers for handling AEs with various perturbation strengths. The LP-Defense layer SAEC and the SP-Defense layer ContraNet in MixDefense are complementary to each other, which can filter out AEs with large and small perturbations in tandem. 

In general, our MixDefense yields superior results among a wide range of perturbation budgets, though other competitors might be more good at detecting AEs of a certain perturbation strength. 
For example,  

MagNet is on par with MixDefense on MNIST once the adversarial perturbation budget of the $L_{2}$ norm attack becomes larger than $0.07$. However, this leading position will vanish when looking at the results on CIFAR10/GTSRB dataset, or when the perturbation budget is small. 
Besides, the $Acc_{\text{Detector}}$ of MixDefense under various perturbation budgets is more stable than the other two AE detectors. 
For instance, there exists a significant positive correlation between the $Acc_{\text{Detector}}$ of FS and the perturbation budget under BIM attack. 

\begin{figure*}[t]
    \centering
    \includegraphics[width=\linewidth]{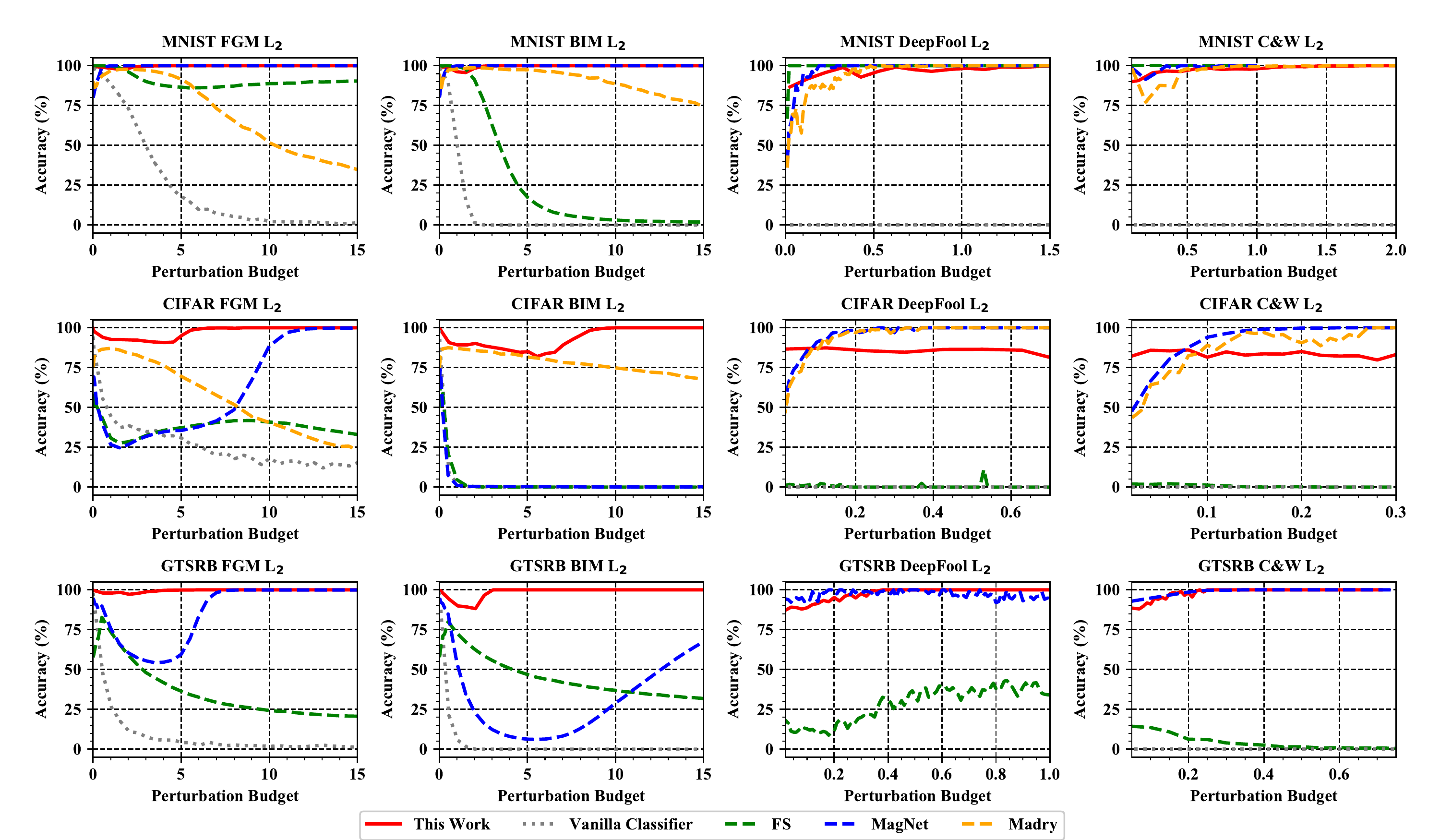}
    \caption{ $Acc_{\text{RC}}$ \textit{vs.} perturbation budget curves under $L_{2}$ norm. The column presents attack method, and the row shows their dataset). The gray dash line refers to the classifier's prediction accuracy(without defense) while the other four colorful lines display the robust accuracy(classifier+defense) curves, where the red solid line indicates the {\itshape{MixDefense}}, green dash line refers to {\itshape{FS}}, blue dash one demonstrates {\itshape{MagNet}} and the yellow dash line displays {\itshape{Madry}}.}
    \label{fig:classifierACCL2}
\end{figure*}

\noindent
{\bfseries Accuracy of Robust Classifier.} 
We regard the classifier supported by the detector as a robust classifier, and report its accuracy in Figure \ref{fig:classifierACCL2}.
We also provide the accuracy of the classifier against attacks without any defense using the dot line colored by gray. 
As shown in Figure~\ref{fig:classifierACCL2}, the accuracy of the vanilla classifier decreases significantly as the perturbation gradually grows. 
Madry also shows a decline trend on CIFAR10 and MNIST but the performance drop is less severe compared with the vanilla classifier. 
There may be some cases where the other two competitors achieve better performance. However, the $Acc_{\text{RC}}$ of MixDefense is more stable compared with FS and MagNet, and can keep a high accuracy value across various attack types or datasets generally. 
For example, Magnet outperforms Mixdefense against DeepFool attack on CIFAR10 when the perturbation larger than 0.1, whereas achieving only 75\% accuracy under the same perturbation budget when it comes to BIM attack. Similarly, FS gains excellent performance on MNIST against FGM attak, while a notable decline could be observed when switching to CIFAR10 and GTSRB. 

As pointed out in Section~\ref{sec:introduction}, an ideal defense strategy should not cause accuracy loss for normal inputs. In view of this, we further test our MixDefense on clean image data, and achieve respective 98.2\%, 90.3\%, and 96.8\% classification accuracy on MNIST, CIFAR10 and GTSRB, which are on par with the original accuracy of vanilla classifier as list in Table~\ref{tab:DataSets}. However, other AE detectors would induce accuracy reduction on clean images. For example,  Madry leads to 7.2\% and 0.63\% accuracy decrease on CIFAR10 and MNIST, respectively.

\subsection{Results of LP-Defense}
\label{sec:exp_lp}

\begin{figure*}[htbp]
    \centering
    \includegraphics[width=\linewidth]{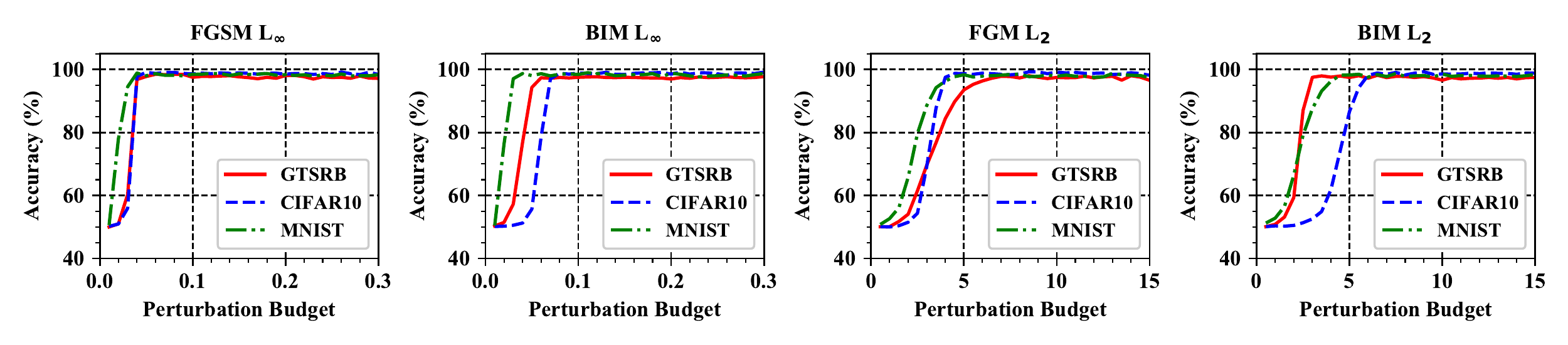}
    \caption{Large perturbation detector(SAEC)'s accuracy v.s. perturbation budget.}
    \label{fig:FINALlarge}
\end{figure*}

\begin{figure}[t]
    \centering
    \includegraphics[width=\linewidth]{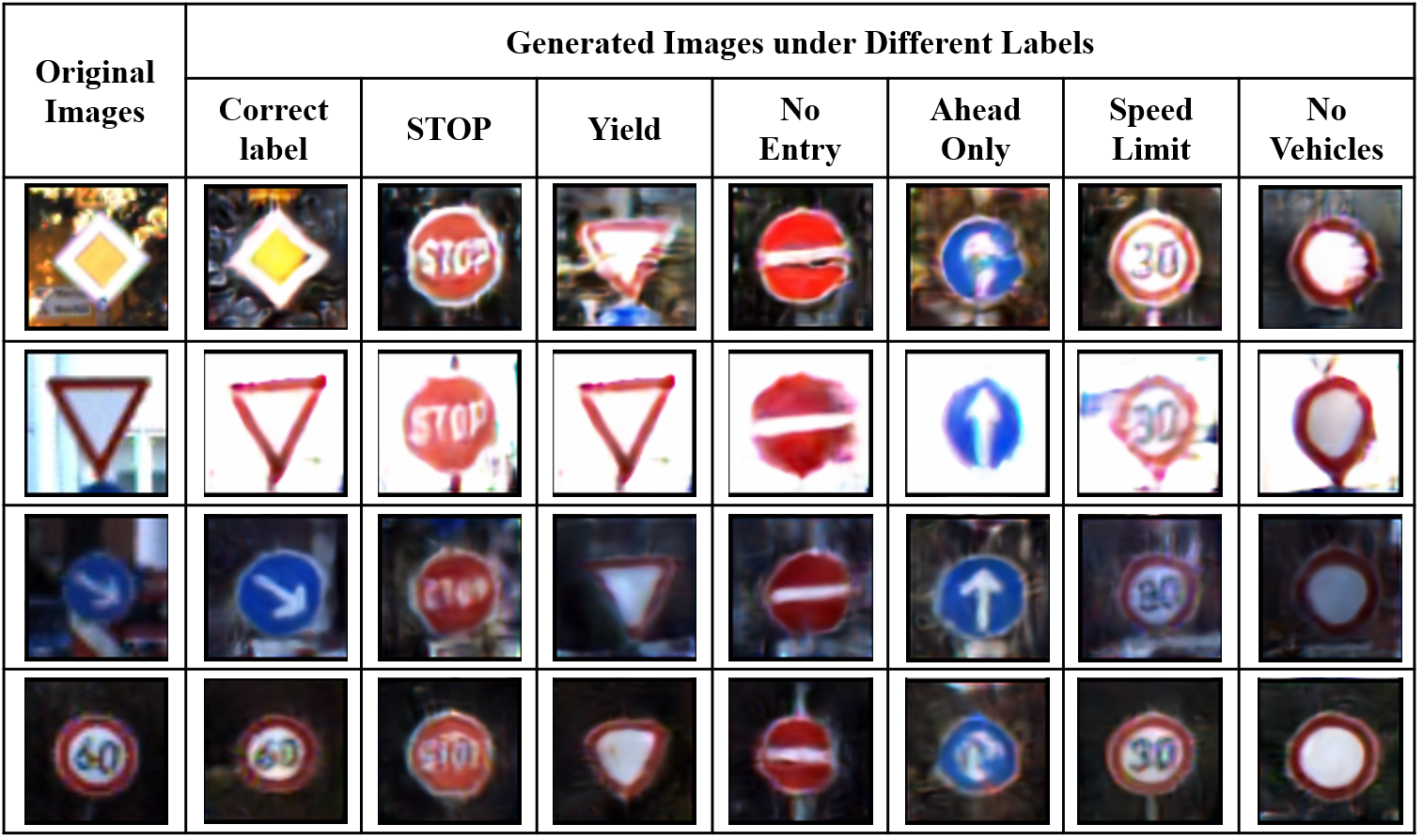}
    \caption{ The correlation of generated image $\mathbf{x}'$ with the conditional label $y$. In each row, the first element is the input image $\mathbf{x}$, the second element depicts its  reconstruction generated using correct label, following reconstructions under other 6 conditional label respectively.}
    \label{fig:GTSRB}
\end{figure}

The performance of SAEC relies on selecting a reasonable threshold for the NL\_scores to distinguish AEs from the normal samples correctly. 
In practice, we use the training data perturbed by random uniform noise instead of the real AEs, and select a proper threshold such that clean images would not be mistakenly regarded as perturbed ones. In this way, we avoid the need of extra information about AEs and complicated optimization, making SAEC easier to be generalized to a wide range of AEs.

Figure \ref{fig:FINALlarge} shows the effectiveness of SAEC by depicting the detector's accuracy $Acc_{\text{Detector}}$ \textit{v.s.} perturbation budget $\epsilon$ against 4 attacks across three datasets. The first two panels demonstrate the detection accuracy against FGSM and BIM, under $L_{\infty}$, and the last two panels are under $L_{2}$ norm. It can be clearly seen in the last two graphs that the accuracy grow dramatically as the perturbation becoming larger. More specifically, once the adversarial perturbation is larger than 0.07, the accuracy will reach a peak plateau in the number above 97\% and will remain steady for larger perturbation. This trend is the same as what we have analysis in previous section. So, if we regard the perturbation budget larger than 0.07 as large perturbation, our SAEC can achieve at least 97\% accuracy against all three attacks in a large perturbation range. When it comes to the $L_{\infty}$ norm attack,  the first two graphs of Figure \ref{fig:FINALlarge} reveal the similar trends, i.e. the accuracy sharply rises to a high point, at least 97\%, when the perturbation above 7, then remains steady.  Taken together, these results indicate that our SAEC is an effective sample statistics method to detect AEs with large perturbation.

\subsection{Results of SP-Defense}
\label{sec:exp_sp}

\noindent
{\bfseries Qualitative results.} 
Figure~\ref{fig:GTSRB} shows the reconstructed images of the revised cGAN in the proposed SP-Defense layer, ContraNet, 

where the first column plots four random sampled original images from the test set directly, the second column represents the reconstruction images of these four images given the correct labels as the conditional information. In contrast to the reconstruction with correct labels, we depict the reconstruction image with another six fixed labels (may or may not equal to their correct label), as shown in the following columns respectively. These figures are a good illustration to show the effectiveness of our small perturbation detection's core idea -- contradiction between the AE's prediction label and its semantic information. More reconstruction examples could be found in Appendix.

\begin{figure*}[htbp]
    \centering
    \includegraphics[width=\linewidth]{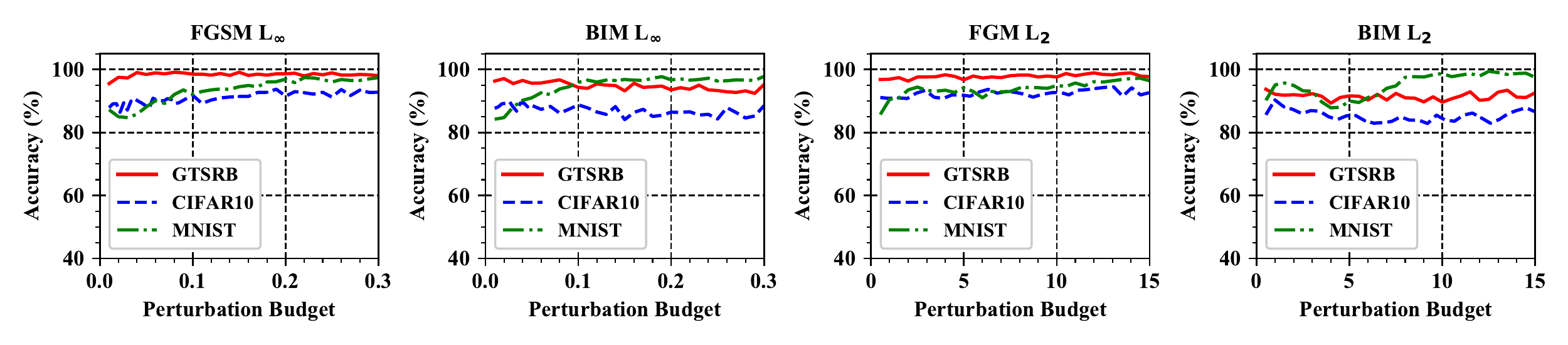}
    \caption{Small perturbation detector's accuracy v.s. perturbation budget. We test ContraNet's performance against FGSM, FGM and BIM under $L_{2}$ or $L_{\infty}$ norm, across different dataset indicated by color. }
    \label{fig:FINALsmall}
\end{figure*}

\noindent
{\bfseries Quantitative results.} 
Figure \ref{fig:FINALsmall} demonstrates our small perturbation detector's accuracy concerning perturbation under the same setting as the large perturbation. As shown in the first two graphs of Figure \ref{fig:FINALsmall}, the detector's accuracy remains steady at a reasonable high level, i.e., at least 85\%, in contrast to the SAEC's poor behavior when the adversarial perturbation is a relatively small range, i.e. 0.01  to 0.07 or even wider range such as 0.01 to 0.1. Consistent with our previous analysis, the small perturbation detector plays a complement role for SAEC, and these two defense layers achieve high accuracy against wide range of adversarial perturbations. This finding also accords with the results shown in the last two graphs of Figure \ref{fig:FINALsmall}, where $L_2$ norm acts as the distance measure. In contrast to our SAEC's deteriorative accuracy against small perturbations, i.e., 0.5 to 7, the small perturbation detector can achieve a stable accuracy, averaging above 90\%,  against different attack methods across all three datasets. The result for MNIST is  counterintuitive, since MNIST is simpler than the other two datasets, while the performance seems poor especially when the perturbation is extremely small. A possible explanation for this might be that the perturbation is too small to obtain enough successful AE sample leading to an unstable statistics result.

\begin{figure}[t]
    \centering
    \includegraphics[width=\linewidth]{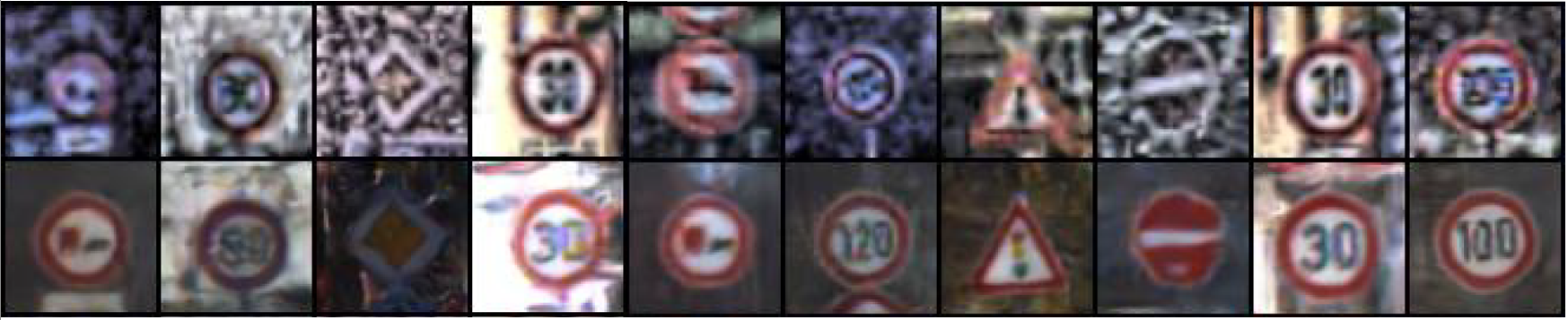}
    \caption{Failure cases of Mixdefense. There lists 10 clean images in the first line which are wrongly judged as AE by Mixdefense. Their corresponding reconstruction images are following in the second line.}
    \label{fig:failurecaseFN}
\end{figure}

\begin{figure}[t]
    \centering
    \includegraphics[width=\linewidth]{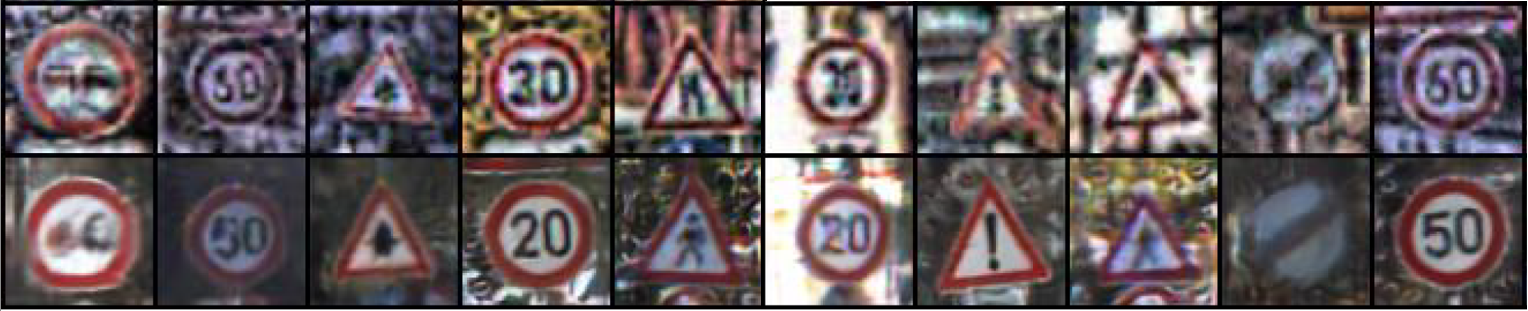}
    \caption{Failure cases of Mixdefense. There lists 10 AE generated by C\&W attack in the first line which are wrongly judged as clean images by Mixdefense. Their corresponding reconstruction images are following in the second line.}
    \label{fig:failurecaseFP}
\end{figure}

\noindent
{\bfseries{Failure case analysis.}} Two batch of failure cases are exemplified in Figure~\ref{fig:failurecaseFN} and Figure~\ref{fig:failurecaseFP} with respect to FN and FP, respectively. It is apparent from both Figures that most input images are corrupted in some degree that it is even a tough work for humans to recognize the image within a limited time. From Figure~\ref{fig:failurecaseFP} we can see that if the adversarial perturbation disturbs the key feature of an image, it will change the semantic meaning of the image and leading the mistake of our ContraNet such as the 6th image pair, whose AE is very similar to its target class that explains why the detection method believe it is a clean image.

\section{Case Study Adaptive Attack}
\label{sec:adaptive_attack}

In the previous section, we show experimental results under \textit{white-box attacks} (Section~\ref{sec:ae_attacks}), wherein attackers have full knowledge of the target classifier but do not consider the proposed MixDefense when generating AEs. To further evaluate the robustness of MixDefense on adaptive attacks, in this section, we present a case study on the GTSRB dataset, which is widely used in the training of autonomous vehicle systems.

\noindent \textbf{Possible attack strategies.}

Perhaps the most straightforward attack strategy is to treat MixDefense and the target classifier as a whole and generate AEs according to the gradient direction of its loss function. However, such a strategy is not practical, because: i) the LP-Defense layer SAEC is not trainable, and ii) the SP-Defense layer ContraNet is not trained in an end-to-end manner. In particular, the target classifier and ContraNet (including the revised cGAN and the deep metric module) have separate gradient propagation processes, and the only connection in between (i.e., the predicted label $y$) is produced by a non-differentiable \texttt{argmax} operator. 

Then, another possible strategy is to attack MixDefense and the target classifier separately and generate a composite AE that combines the two independently generated perturbations. Even though such composition would affect each other's attack strength and hence difficult to perform, we show that it is extremely difficult, if not impossible, to attack MixDefense 
alone in the following.

\noindent \textbf{Robustness of MixDefense.}
Here, we focus on the robustness of ContraNet. This is because, as the last defense layer of MixDefense, as long as ContraNet is robust against adaptive attacks, we can safely conclude MixDefense is also robust. 

\begin{figure}[t]
    \centering
    \includegraphics[width=\linewidth]{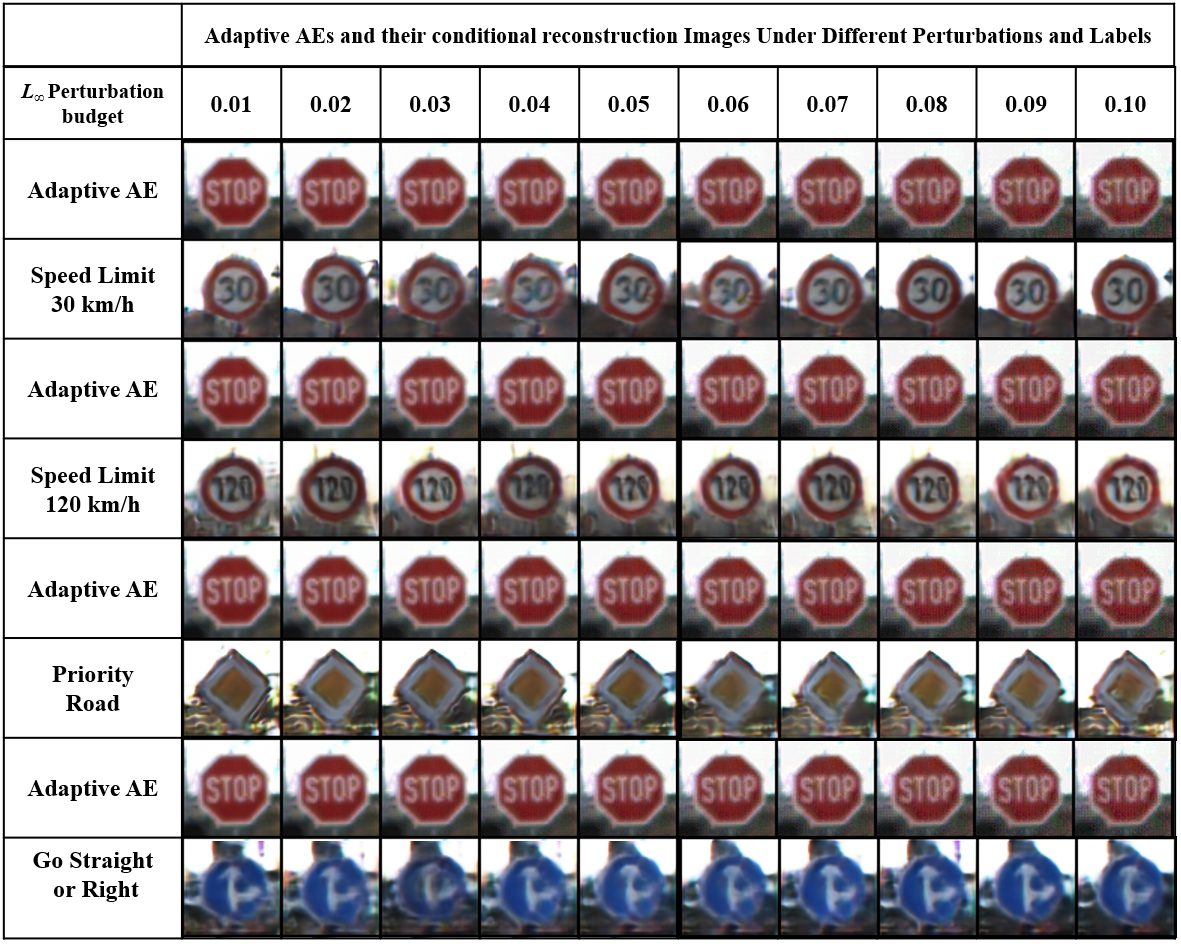}
    \caption{Adaptive attack on the input images of the improved cGAN in the SP-Defense layer of MixDefense with gradually increased perturbation strength. As can be observed, the reconstructed images produced by the improved cGAN are still strongly correlated with the target labels.}
    \label{fig:defenseawareattack2}
\end{figure}

The revised cGAN in ContraNet can generate reconstruction results that are strongly coupled with the conditions, i.e., the predicted labels provided by the target classifier. 
We empirically verify this claim by performing targeted FGSM $L_{\infty}$ attack the inputs of cGAN following~\cite{advonvae}, where the perturbation strength ranges from $0.01$ to $0.1$. For each target attack, we generate AEs under a given perturbation budget in such a manner that the reconstructed image is optimized to be similar to the corresponding AE whenever possible. As can be observed in Figure \ref{fig:defenseawareattack2}, however, due to the strong correlation between the reconstruction and the predicted label, the semantics of the reconstructed images still follow the predicted label under such adaptive attacks, resulting in contradictions that can be exploited for AE detection.  

\begin{figure}[t]
    \centering
    \includegraphics[width=\linewidth]{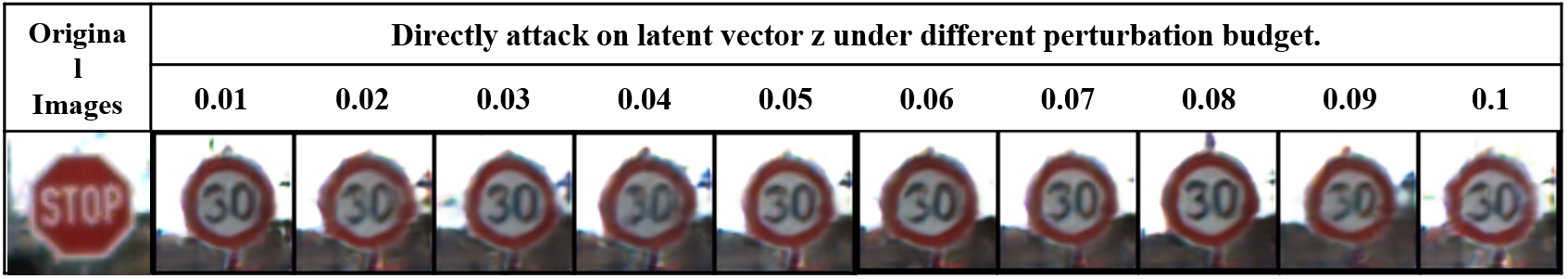}
    \caption{Directly performing adaptive attack on the latent vectors of the improved cGAN. Although we give more advantage to the attackers, i.e. empowering them with the ability to modify the latent vector $\mathbf{z}$ without constrain, the generated images are still strong coupled with the condition ``Speed limit (30km/h)''. See Section~\ref{sec:adaptive_attack} for details.}

    \label{fig:defenseawareattack3}
\end{figure}

Next, let us assume that attackers can bypass the encoder part in ContraNet and attack the revised cGAN by modifying the latent variable $\mathbf{z}$ directly\footnote{In practice, this is not possible because AE attacks can only modify the inputs.}. Under such relaxed attack constraints, the resulted latent variable might not be normally distributed. We fix the class condition $y$ whose semantics is different from the AE, and seek for a $\mathbf{z}$ that can yield a reconstructed image that is similar to the AE. As shown in Figure \ref{fig:defenseawareattack3}, although the strength $\epsilon$ ($L_{\infty}$) of FGSM attack exerted on $\mathbf{z}$ is increased from $0.05$ to $0.3$, the obtained $\mathbf{z}$ could only be reconstructed to `Speed limited 30 km/h' given the condition $y="limited 30 km/h"$. 
Note that, directly attacking $\mathbf{z}$ can be viewed as the strongest possible adaptive attack for ContraNet, because it offers the maximum freedom to affect the revised cGAN by exerting no restrictions on the latent variable $\mathbf{z}$. 

In summary, it is rather difficult, if not impossible, to successfully perform an adaptive attack on MixDefense wherein the attacker has full knowledge of the target model and the defense framework.

\section{Discussion}

\noindent
{\bfseries{Good side effects.}} Deep classifiers would inevitably misclassify some inputs when tested on natural corruption data\cite{ford2019adversarial}. Although no intentional perturbations are exerted on these samples, our MixDefense would still identify these misclassified samples as AEs. This is because there exist contradictions between the semantic meaning of the samples and the predicted labels, while the SP-defense layer ConraNet is designed to detect AEs with small perturbations based on such contradictions. See Section~\ref{sec:SP_defense} for more details. 
Though wrongly recognizing these misclassified samples as AEs, rejecting such samples would not decrease the accuracy of the target classifier and may even improve the classification accuracy~\cite{smith2011improving}.

\noindent
{\bfseries{Intermediate layers.}} 
As mentioned above, our proposed Mixdefense framework consists of multiple defense layers targeted on AEs with various strengths of perturbations. Besides the LP-Defense layer and the SP-defense layer introduced in this paper, one might also deploy one or several intermediate defense layers for those AEs with medium-sized perturbations. This can not only reduce the detection burden offloaded to the last SP-Defense layer but also simplify its design. Existing AE detection techniques reviewed in Section~\ref{sec:ae_defenses} can be adjusted and incorporated as the intermediate defense layers in MixDefense to further bootstrap the defense performance.

\noindent
{\bfseries{Limitations.}} We found that the Contranet can achieve good performance on data sets with obvious category information, such as GTSRB. When the class information of the dataset is relatively vague, such as imageNet, on which the state-of-the-art classifier, {\it NFNet-F6 w/SAM} can achieve 86.5\% accuracy without extra training data\cite{sota}, and some images in the data set can be classified to multiple labels or even with wrong labels
\cite{northcutt2021pervasive}, the performance of ContraNet will  tend to decrease. To improve the performance of ContraNet on complex datasets such as imagenet, an advanced cGAN or more training samples and training time are necessary.







\section{Conclusion}
In this work, we propose \textit{MixDefense}, a multi-layer attack-agnostic defense-in-depth framework for AE detection. For AEs with large perturbations, the proposed SAEC technique in the LP-defense layer is used to discover the \textit{statistical} `noise' difference between natural images and tampered ones. For AEs with small perturbations, the proposed ContraNet in the SP-defense layer effectively captures the contradiction between the inference results of AEs and their \textit{semantic} information. Experimental results with various AE attack methods on image classification datasets show that MixDefense outperforms existing AE detection techniques by a considerable margin and it is also resistant to adaptive attacks.
\label{sec:conclusion}

\bibliographystyle{ieeetr}

\bibliography{ref}

\begin{thebibliography}{10}

\bibitem{szegedy2013intriguing}
C.~Szegedy, W.~Zaremba, I.~Sutskever, J.~Bruna, D.~Erhan, I.~Goodfellow, and
  R.~Fergus, ``Intriguing properties of neural networks,'' in {\em
  International Conference on Learning Representations (ICLR)}, 2014.

\bibitem{silva2020opportunities}
S.~H. Silva and P.~Najafirad, ``Opportunities and challenges in deep learning
  adversarial robustness: A survey,'' {\em arXiv preprint arXiv:2007.00753},
  2020.

\bibitem{madry2017towards}
A.~Madry, A.~Makelov, L.~Schmidt, D.~Tsipras, and A.~Vladu, ``Towards deep
  learning models resistant to adversarial attacks,'' in {\em International
  Conference on Learning Representations (ICLR)}, 2018.

\bibitem{pmlr-v97-pang19a}
T.~Pang, K.~Xu, C.~Du, N.~Chen, and J.~Zhu, ``Improving adversarial robustness
  via promoting ensemble diversity,'' in {\em International Conference on
  Machine Learning (ICML)}, 2019.

\bibitem{Shan2020GottaCA}
S.~Shan, E.~Wenger, B.~Wang, B.~Li, H.~Zheng, and B.~Zhao, ``Gotta catch'em
  all: Using honeypots to catch adversarial attacks on neural networks,'' {\em
  ACM SIGSAC Conference on Computer and Communications Security (CCS)}, 2020.

\bibitem{Lcuyer2019CertifiedRT}
M.~L{\'e}cuyer, V.~Atlidakis, R.~Geambasu, D.~Hsu, and S.~Jana, ``Certified
  robustness to adversarial examples with differential privacy,'' {\em IEEE
  Symposium on Security and Privacy (SP)}, pp.~656--672, 2019.

\bibitem{pmlr-v119-raghunathan20a}
A.~Raghunathan, S.~M. Xie, F.~Yang, J.~Duchi, and P.~Liang, ``Understanding and
  mitigating the tradeoff between robustness and accuracy,'' in {\em
  International Conference on Machine Learning (ICML)}, 2010.

\bibitem{tramer2020fundamental}
F.~Tram{\`{e}}r, J.~Behrmann, N.~Carlini, N.~Papernot, and J.~Jacobsen,
  ``Fundamental tradeoffs between invariance and sensitivity to adversarial
  perturbations,'' in {\em International Conference on Machine Learning
  (ICML)}, 2020.

\bibitem{meng2017magnet}
D.~Meng and H.~Chen, ``Magnet: a two-pronged defense against adversarial
  examples,'' in {\em ACM SIGSAC Conference on Computer and Communications
  Security (CCS)}, 2017.

\bibitem{samangouei2018defense}
P.~Samangouei, M.~Kabkab, and R.~Chellappa, ``Defense-gan: Protecting
  classifiers against adversarial attacks using generative models,'' in {\em
  International Conference on Learning Representations (ICLR)}, 2018.

\bibitem{Dziugaite2016ASO}
G.~Dziugaite, Z.~Ghahramani, and D.~M. Roy, ``A study of the effect of jpg
  compression on adversarial images,'' {\em ArXiv}, vol.~abs/1608.00853, 2016.

\bibitem{Grosse2017OnT}
K.~Grosse, P.~Manoharan, N.~Papernot, M.~Backes, and P.~McDaniel, ``On the
  (statistical) detection of adversarial examples,'' {\em ArXiv}, 2017.

\bibitem{Song2018PixelDefendLG}
Y.~Song, T.~Kim, S.~Nowozin, S.~Ermon, and N.~Kushman, ``Pixeldefend:
  Leveraging generative models to understand and defend against adversarial
  examples,'' in {\em International Conference on Learning Representations
  (ICLR)}, 2018.

\bibitem{Aigrain2019DetectingAE}
J.~Aigrain and M.~Detyniecki, ``Detecting adversarial examples and other
  misclassifications in neural networks by introspection,'' {\em ArXiv},
  vol.~abs/1905.09186, 2019.

\bibitem{Carrara2018AdversarialED}
F.~Carrara, R.~Becarelli, R.~Caldelli, F.~Falchi, and G.~Amato, ``Adversarial
  examples detection in features distance spaces,'' in {\em European Conference
  on Computer Vision (ECCV Workshops)}, 2018.

\bibitem{ma2018characterizing}
X.~Ma, B.~Li, Y.~Wang, S.~M. Erfani, S.~Wijewickrema, G.~Schoenebeck, D.~Song,
  M.~E. Houle, and J.~Bailey, ``Characterizing adversarial subspaces using
  local intrinsic dimensionality,'' in {\em International Conference on
  Learning Representations (ICLR)}, 2018.

\bibitem{kantaros2020visionguard}
Y.~Kantaros, T.~Carpenter, S.~Park, R.~Ivanov, S.~Jang, I.~Lee, and J.~Weimer,
  ``Visionguard: Runtime detection of adversarial inputs to perception
  systems,'' {\em arXiv preprint arXiv:2002.09792}, 2020.

\bibitem{xu2018feature}
W.~Xu, D.~Evans, and Y.~Qi, ``Feature squeezing: Detecting adversarial examples
  in deep neural networks,'' in {\em The Network and Distributed System
  Security Symposium (NDSS)}, 2018.

\bibitem{tramer2018ensemble}
F.~Tram{\`e}r, A.~Kurakin, N.~Papernot, I.~Goodfellow, D.~Boneh, and
  P.~McDaniel, ``Ensemble adversarial training: Attacks and defenses,'' in {\em
  International Conference on Learning Representations (ICLR)}, 2018.

\bibitem{Cheval2018DEEPSECDE}
V.~Cheval, S.~Kremer, and I.~Rakotonirina, ``Deepsec: Deciding equivalence
  properties in security protocols theory and practice,'' {\em IEEE Symposium
  on Security and Privacy (SP)}, 2018.

\bibitem{fgsm}
I.~J. Goodfellow, J.~Shlens, and C.~Szegedy, ``Explaining and harnessing
  adversarial examples,'' in {\em International Conference on Learning
  Representations (ICLR)}, 2015.

\bibitem{kurakin2016bim}
A.~Kurakin, I.~J. Goodfellow, and S.~Bengio, ``Adversarial examples in the
  physical world,'' in {\em International Conference on Learning
  Representations (ICLR workshop)}, 2017.

\bibitem{papernot2016limitations}
N.~Papernot, P.~McDaniel, S.~Jha, M.~Fredrikson, Z.~B. Celik, and A.~Swami,
  ``The limitations of deep learning in adversarial settings,'' in {\em IEEE
  European Symposium on Security and Privacy (EuroSP)}, 2016.

\bibitem{Pang2020ATO}
R.~Pang, H.~Shen, X.~Zhang, S.~Ji, Y.~Vorobeychik, X.~Luo, A.~X. Liu, and
  T.~Wang, ``A tale of evil twins: Adversarial inputs versus poisoned models,''
  {\em ACM SIGSAC Conference on Computer and Communications Security (CCS)},
  2020.

\bibitem{MoosaviDezfooli2016DeepFoolAS}
S.-M. Moosavi-Dezfooli, A.~Fawzi, and P.~Frossard, ``Deepfool: A simple and
  accurate method to fool deep neural networks,'' {\em IEEE Conference on
  Computer Vision and Pattern Recognition (CVPR)}, pp.~2574--2582, 2016.

\bibitem{cw2017}
N.~Carlini and D.~Wagner, ``Towards evaluating the robustness of neural
  networks,'' in {\em IEEE Symposium on Security and Privacy (SP)}, 2017.

\bibitem{xiao2018spatially}
C.~Xiao, J.~Zhu, B.~Li, W.~He, M.~Liu, and D.~Song, ``Spatially transformed
  adversarial examples,'' in {\em International Conference on Learning
  Representations (ICLR)}, 2018.

\bibitem{baluja2017adversarial}
S.~Baluja and I.~Fischer, ``Adversarial transformation networks: Learning to
  generate adversarial examples,'' {\em arXiv preprint arXiv:1703.09387}, 2017.

\bibitem{bulusu2020anomalous}
S.~Bulusu, B.~Kailkhura, B.~Li, P.~K. Varshney, and D.~Song, ``Anomalous
  example detection in deep learning: A survey,'' {\em IEEE Access}, 2020.

\bibitem{kannan2018adversarial}
H.~Kannan, A.~Kurakin, and I.~Goodfellow, ``Adversarial logit pairing,'' 2018.

\bibitem{ziang2018deepdefense}
Z.~Yan, Y.~Guo, and C.~Zhang, ``Deep defense: Training dnns with improved
  adversarial robustness,'' NeurIPS, 2018.

\bibitem{pmlrv97zhang19p}
H.~Zhang, Y.~Yu, J.~Jiao, E.~Xing, L.~E. Ghaoui, and M.~Jordan, ``Theoretically
  principled trade-off between robustness and accuracy,'' in {\em Proceedings
  of the 36th International Conference on Machine Learning}, 2019.

\bibitem{papernot2016distillation}
N.~Papernot, P.~McDaniel, X.~Wu, S.~Jha, and A.~Swami, ``Distillation as a
  defense to adversarial perturbations against deep neural networks,'' in {\em
  IEEE Symposium on Security and Privacy (SP)}, 2016.

\bibitem{Liu2018TowardsRN}
X.~Liu, M.~Cheng, H.~Zhang, and C.~Hsieh, ``Towards robust neural networks via
  random self-ensemble,'' in {\em European Conference on Computer Vision
  (ECCV)}, 2018.

\bibitem{Das2018SHIELDFP}
N.~Das, M.~Shanbhogue, S.-T. Chen, F.~Hohman, S.~Li, L.~Chen, M.~E. Kounavis,
  and D.~H. Chau, ``Shield: Fast, practical defense and vaccination for deep
  learning using jpeg compression,'' {\em ACM SIGKDD International Conference
  on Knowledge Discovery and Data Mining (KDD)}, 2018.

\bibitem{Guo2018CounteringAI}
C.~Guo, M.~Rana, M.~Ciss{\'{e}}, and L.~van~der Maaten, ``Countering
  adversarial images using input transformations,'' in {\em International
  Conference on Learning Representations (ICLR)}, 2018.

\bibitem{Zhang2019DefendingAW}
Y.~Zhang and P.~Liang, ``Defending against whitebox adversarial attacks via
  randomized discretization,'' in {\em International Conference on Artificial
  Intelligence and Statistics {AISTATS}} (K.~Chaudhuri and M.~Sugiyama, eds.),
  2019.

\bibitem{Xie2018MitigatingAE}
C.~Xie, J.~Wang, Z.~Zhang, Z.~Ren, and A.~Yuille, ``Mitigating adversarial
  effects through randomization,'' in {\em International Conference on Learning
  Representations (ICLR)}, 2018.

\bibitem{defense-gan-2018}
S.~Pouya, K.~Maya, and C.~Rama, ``Defense-gan: Protecting classifiers against
  adversarial attacks using generative models,'' in {\em International
  Conference on Learning Representations (ICLR)}, 2018.

\bibitem{Metzen2017OnDA}
J.~H. Metzen, T.~Genewein, V.~Fischer, and B.~Bischoff, ``On detecting
  adversarial perturbations,'' in {\em International Conference on Learning
  Representations (ICLR)}, 2017.

\bibitem{Feinman2017DetectingAS}
R.~Feinman, R.~R. Curtin, S.~Shintre, and A.~B. Gardner, ``Detecting
  adversarial samples from artifacts,'' {\em ArXiv}, vol.~abs/1703.00410, 2017.

\bibitem{SAMP}
J.~Liu, W.~Zhang, Y.~Zhang, D.~Hou, Y.~Liu, H.~Zha, and N.~Yu, ``Detection
  based defense against adversarial examples from the steganalysis point of
  view,'' in {\em IEEE/CVF Conference on Computer Vision and Pattern
  Recognition (CVPR)}, 2019.

\bibitem{Zheng2018RobustDO}
Z.~Zheng and P.~Hong, ``Robust detection of adversarial attacks by modeling the
  intrinsic properties of deep neural networks,'' in {\em NeurIPS}, 2018.

\bibitem{Ma2019NICDA}
S.~Ma, Y.~Liu, G.~Tao, W.~Lee, and X.~Zhang, ``Nic: Detecting adversarial
  samples with neural network invariant checking,'' in {\em The Network and
  Distributed System Security Symposium (NDSS)}, 2019.

\bibitem{Miller2019WhenNT}
D.~Miller, Y.~Wang, and G.~Kesidis, ``When not to classify: Anomaly detection
  of attacks (ada) on dnn classifiers at test time,'' {\em Neural Computation},
  2019.

\bibitem{wang2019adversarial}
J.~Wang, G.~Dong, J.~Sun, X.~Wang, and P.~Zhang, ``Adversarial sample detection
  for deep neural network through model mutation testing,'' in {\em ICSE},
  2019.

\bibitem{liang2021}
B.~Liang, H.~Li, M.~Su, X.~Li, W.~Shi, and X.~Wang, ``Detecting adversarial
  image examples in deep neural networks with adaptive noise reduction,'' {\em
  IEEE Transactions on Dependable and Secure Computing (TDSC)}, 2021.

\bibitem{steg1}
M.~Goljan, J.~Fridrich, and T.~Holotyak, ``New blind steganalysis and its
  implications,'' in {\em Security, Steganography, and Watermarking of
  Multimedia Contents VIII}, 2006.

\bibitem{steg0}
S.~Lyu and H.~Farid, ``Detecting hidden messages using higher-order statistics
  and support vector machines,'' in {\em International Workshop on information
  hiding}, 2002.

\bibitem{steg3}
T.~Pevny, P.~Bas, and J.~Fridrich, ``Steganalysis by subtractive pixel
  adjacency matrix,'' {\em IEEE Transactions on information Forensics and
  Security}, 2010.

\bibitem{steg4}
D.~Zou, Y.~Q. Shi, W.~Su, and G.~Xuan, ``Steganalysis based on markov model of
  thresholded prediction-error image,'' in {\em IEEE International conference
  on multimedia and expo}, 2006.

\bibitem{steg5}
J.~Fridrich and J.~Kodovsky, ``Rich models for steganalysis of digital
  images,'' {\em IEEE Transactions on Information Forensics and Security},
  2012.

\bibitem{liu2019detection}
J.~Liu, W.~Zhang, Y.~Zhang, D.~Hou, Y.~Liu, H.~Zha, and N.~Yu, ``Detection
  based defense against adversarial examples from the steganalysis point of
  view,'' in {\em IEEE/CVF Conference on Computer Vision and Pattern
  Recognition (CVPR)}, 2019.

\bibitem{miyato2018cgans}
T.~Miyato and M.~Koyama, ``cgans with projection discriminator,'' in {\em
  International Conference on Learning Representations (ICLR)}, 2018.

\bibitem{triplet}
E.~Hoffer and N.~Ailon, ``Deep metric learning using triplet network,'' in {\em
  International Workshop on Similarity-Based Pattern Recognition}, 2015.

\bibitem{tripletloss}
K.~Q. Weinberger and L.~K. Saul, ``Distance metric learning for large margin
  nearest neighbor classification.,'' {\em Journal of Machine Learning
  Research}, 2009.

\bibitem{miner}
F.~Schroff, D.~Kalenichenko, and J.~Philbin, ``Facenet: A unified embedding for
  face recognition and clustering,'' in {\em IEEE conference on computer vision
  and pattern recognition (CVPR)}, 2015.

\bibitem{metric0}
K.~Musgrave, S.~Belongie, and S.-N. Lim, ``Pytorch metric learning,'' 2020.

\bibitem{mnist}
Y.~LeCun, C.~Cortes, and C.~Burges, ``The mnist dataset,'' {\em
  http://yann.lecun.com/exdb/mnist/}.

\bibitem{krizhevsky2014cifar}
A.~Krizhevsky, V.~Nair, and G.~Hinton, ``The cifar-10 dataset,'' {\em online:
  http://www. cs. toronto. edu/kriz/cifar. html}, 2014.

\bibitem{GTSRB}
J.~Stallkamp, M.~Schlipsing, J.~Salmen, and C.~Igel, ``Man vs. computer:
  Benchmarking machine learning algorithms for traffic sign recognition,'' {\em
  Neural Networks}.

\bibitem{kos2018adversarial}
J.~Kos, I.~Fischer, and D.~Song, ``Adversarial examples for generative
  models,'' in {\em IEEE SPW}, pp.~36--42, 2018.

\bibitem{carlini2019evaluating}
N.~Carlini, A.~Athalye, N.~Papernot, W.~Brendel, J.~Rauber, D.~Tsipras,
  I.~Goodfellow, A.~Madry, and A.~Kurakin, ``On evaluating adversarial
  robustness,'' {\em arXiv preprint arXiv:1902.06705}, 2019.

\bibitem{deepsec}
X.~Ling, S.~Ji, J.~Zou, J.~Wang, C.~Wu, B.~Li, and T.~Wang, ``Deepsec: A
  uniform platform for security analysis of deep learning model,'' in {\em IEEE
  Symposium on Security and Privacy (SP)}, 2019.

\bibitem{foolbox}
J.~Rauber, R.~Zimmermann, M.~Bethge, and W.~Brendel, ``Foolbox native: Fast
  adversarial attacks to benchmark the robustness of machine learning models in
  pytorch, tensorflow, and jax,'' {\em Journal of Open Source Software},
  vol.~5, no.~53, p.~2607, 2020.

\bibitem{cleverhans}
N.~Papernot, F.~Faghri, N.~Carlini, I.~Goodfellow, R.~Feinman, A.~Kurakin,
  C.~Xie, Y.~Sharma, T.~Brown, A.~Roy, A.~Matyasko, V.~Behzadan,
  K.~Hambardzumyan, Z.~Zhang, Y.-L. Juang, Z.~Li, R.~Sheatsley, A.~Garg,
  J.~Uesato, W.~Gierke, Y.~Dong, D.~Berthelot, P.~Hendricks, J.~Rauber, and
  R.~Long, ``Technical report on the cleverhans v2.1.0 adversarial examples
  library,'' {\em arXiv preprint arXiv:1610.00768}, 2018.

\bibitem{moosavi2016deepfool}
S.-M. Moosavi-Dezfooli, A.~Fawzi, and P.~Frossard, ``Deepfool: a simple and
  accurate method to fool deep neural networks,'' in {\em IEEE/CVF Conference
  on Computer Vision and Pattern Recognition (CVPR)}, 2016.

\bibitem{advonvae}
G.~Gondim-Ribeiro, P.~Tabacof, and E.~Valle, ``Adversarial attacks on
  variational autoencoders,'' {\em arXiv preprint arXiv:1806.04646}, 2018.

\bibitem{ford2019adversarial}
N.~Ford, J.~Gilmer, N.~Carlini, and D.~Cubuk, ``Adversarial examples are a
  natural consequence of test error in noise,'' 2019.

\bibitem{smith2011improving}
M.~R. Smith and T.~Martinez, ``Improving classification accuracy by identifying
  and removing instances that should be misclassified,'' in {\em International
  Joint Conference on Neural Networks (IJCNN)}, pp.~2690--2697, 2011.

\bibitem{sota}
``Leader board of image classification on imagenet.''
  \url{https://paperswithcode.com/sota/image-classification-on-imagenet}, 2021.

\bibitem{northcutt2021pervasive}
C.~G. Northcutt, A.~Athalye, and J.~Mueller, ``Pervasive label errors in test
  sets destabilize machine learning benchmarks,'' 2021.

\end{thebibliography}



   

   



\end{document}